\documentclass[journal]{IEEEtran}
\usepackage{bm}
\usepackage{graphicx}
\usepackage{subfigure}
\usepackage{graphics}
\usepackage{algorithm}
\usepackage{algorithmic}
\usepackage{epstopdf}
\usepackage{color}
\usepackage{amsmath}
\usepackage{indentfirst}
\usepackage{cite}
\usepackage{url}
\usepackage{setspace}
\usepackage{booktabs}
\usepackage{multirow}
\usepackage{amssymb}
\usepackage{times}
\usepackage{stfloats}
\usepackage{enumitem}
\usepackage{pifont}

\ifCLASSINFOpdf
\else
\fi
\begin{document}
\title{Secure Long-range Autonomous Valet Parking: A Reservation Scheme with Three-factor Authentication and Key Agreement}
\author{Di~Wang,
        Yue~Cao,~\IEEEmembership{Senior Member,~IEEE}, Fei Yan,~\IEEEmembership{Member,~IEEE}, Yining Liu,~\IEEEmembership{Member,~IEEE}, Daxin Tian,~\IEEEmembership{Senior Member,~IEEE}, and Yuan Zhuang, ~\IEEEmembership{Member,~IEEE}
        % <-this % stops a space
\thanks{D. Wang, Y. Cao (Corresponding author), and F. Yan are with the School of Cyber Science and Engineering, Wuhan University, Wuhan 430072, China (e-mail:1065889068@qq.com; yue.cao@whu.edu.cn; yanfei@whu.edu.cn).}
\thanks{Y. Liu is with the Institute for Artificial Intelligence Interdisciplinary Research, Guilin University of Electronic Technology Guilin 541004, China (e-mail:ynliu@guet.edu.cn).}
\thanks{D. Tian is with the School of Transportation Science and Engineering, Beihang University, Beijing 100191, China (e-mail:dtian@buaa.edu.cn).}
\thanks{Y. Zhuang is with the State Key Laboratory of Surveying, Mapping and Remote Sensing, Wuhan University, Wuhan 430072, China (e-mail: yuan.zhuang@whu.edu.cn).}}
%{Shell \MakeLowercase{\textit{et al.}}: Bare Demo of IEEEtran.cls for IEEE Journals}
\maketitle

\begin{abstract}
Long-range autonomous valet parking (LAVP) is a current trend, partly due to traffic congestion and parking headache. For large-scale parking demands, reservation is introduced to effectively manage valet parking. However, existing schemes focus on parking request verification and parking check-in, which aren't applicable LAVP because they ignore identity legitimacy and communication security in the phase of picking up as well as dropping off passengers. One viable solution is authentication and key agreement (AKA) protocol. Generally, due to low entropy of passwords and dictionary attacks, three-factor (i.e. passwords, biometrics, and smart card) AKA is more secure than single- and two-factor AKA. Unfortunately, known attacks and high overheads hinder the application of three-factor AKA in real-world environments. Hence, one of the most tough tasks is to balance security and availability, especially how to address the potential threats introduced by each factor while taking full advantage of three factors. Inspired by the above challenges, we propose a provably secure three-factor AKA protocol for reservation services in LAVP, namely SecLAVP. Specifically, the passenger and the autonomous vehicle (\bm{$AV$}) complete mutual authentication with the assistance of the drop-off/pick-up point (\bm{$DP$}). After successful authentication, the session key is generated between the passenger, \bm{$DP$}, and \bm{$AV$} for secure communication. In the Real-Or-Random (ROR) model, we formally prove SecLAVP satisfies session-key security. Furthermore, we apply AVISPA to simulate our proposed protocol, to demonstrate that SecLAVP can resist man-in-the-middle attacks. Additionally, informal security analysis indicates that SecLAVP satisfies our defined 15 design goals concerning security. Finally, we evaluate performance of SecLAVP in terms of communication overheads, computational overheads, and scheduling, to manifest the feasibility.
\end{abstract}
\vspace{-1mm}

\IEEEpeerreviewmaketitle

\section{Introduction}
\IEEEPARstart{P}{arking} issues have become a commonplace and frustrating conundrum in urban life \cite{1}, partly attested that 30$\%$ of traffic jams are triggered by cruising for a parking space. More importantly, searching for an idle parking space causes air pollution \cite{2}. For example, in 2006, searching for a vacant parking space gives rise to travel 945,000 additional miles, consume 47,000 gallons of gasoline and generate 728 tons of air pollution in Los Angeles \cite{3}.

As one of the foremost applications in the field of autonomous driving, autonomous valet parking (AVP) is divided into short-range autonomous valet parking (SAVP) \cite{4,5,6,7} and long-range autonomous valet parking (LAVP) \cite{8,9,10,11}. Daimler-Benz \cite{12} took the lead in SAVP, which is semi-autonomous parking since passengers need park autonomous vehicles ($AVs$) at the entrance of the parking lot. Subsequently, ZongMu \cite{13} deployed LAVP, which is fully autonomous parking since passengers only park $AVs$ at a designated location (i.e. drop-off/pick-up points, $DPs$) near the destination. To improves last-kilometer experience, we focus on LAVP.

Although navigation systems (e.g., BeiDou Navigation Satellite System and Baidu Map) can guide to parking lots near the destination, the parking space is still occupied after $AVs$ arriving during rush hours. Additionally, for large-scale parking demands, the standard service of LAVP (e.g., first-come-first-serve) causes traffic congestion and long waiting time, along with poor user experience and quality of service.

The above challenges highlight the significance of proposing a reservation scheme to effectively manage valet parking. When proposing such a reservation, the relevant security and privacy requirements must be a top priority. Reservations sent by $AVs$ to third parties include sensitive information (e.g., parking duration, identity and location). Once such information is captured, attackers can construct characteristics concerning passenger, such as work place and home address. Furthermore, situation is even grim as terrorists launch terrorist incidents based on the location and trajectory of $AVs$. Finally, service providers may sell users' private message to others for increasing profits. Therefore, references \cite{14,15,16,17,18} paid attention to location and trajectory privacy protection of SAVP.

The above schemes ignore communication security, that is, confidentiality, unforgeability, and access control. First, attackers are easy to breach confidentiality by eavesdropping message in an open wireless environment. Next, tampered or deleted reservations will affect the accuracy of reservation decisions and user experience. Finally, illegal access exacerbates threats. For instance, attackers exploiting existing vulnerabilities to remotely control $AVs$ to launch coordinated attacks.

The top priority is to propose an authentication and key agreement (AKA) scheme to secure communication. Initially, researchers focused on single factor AKA \cite{19,20} and two-factor AKA \cite{21,22,23}, facing issues such as smart card loss, passwords guessing, and biometrics leakage. Subsequently, three-factor (smart card, passwords, and biometrics) AKA \cite{24,25,26,27,28,29,42,30,31} has attracted widespread attention. However, AKA that integrates three factors inevitably introduces threats faced by each factor. Therefore, the most tough task is to address the potential threats of each factor while taking full advantage of three factors.

Motivated by the above challenge, we propose a secure reservation scheme for LAVP with three-factor authentication and key agreement (hereinafter referred to as SecLAVP), based on our prior parking optimization \cite{10}. The main contributions of this paper are as follow:
\begin{enumerate}
\item To prevent $AVs$ from being remotely controlled, unauthorized data access, and data eavesdropping, tampering, or forgery, we propose a three-factor AKA protocol based on smart card, passwords, and biometrics (i.e. SecLAVP). SecLAVP realizes authentication among passengers, $DPs$ as well as $AVs$, and then session key is generated for secure communication. More specifically, SecLAVP supports passwords and biometrics update. No passwords verified table and smart card revocation are utilized to resist passwords guessing and smart card loss, respectively. Additionally, biometric extraction is applied to prevent biometrics leakage.

\item To further enhance security, we have conducted formal security proof and formal security verification. On the one hand, under the widely accepted Real-Or-Random (ROR) model \cite{32}, we prove that SecLAVP satisfies semantic security based on elliptic curve decisional Diffie Hellman ($ECDDH$) problem. On the other hand, we verify that SecLAVP resists man-in-the-middle attacks via Automated Validation of Internet Security Protocol and Applications (AVISPA) \cite{33}.

\item Extensive simulations are carried out. First, we analyze the security to indicate that SecLAVP not only satisfies anonymity and forward security, but also resists common attacks such as offline dictionary attacks and replay attacks. Then, we evaluate efficiency of SecLAVP in term of computational overheads and communication overheads. Finally, to validate the practicability of SecLAVP, we evaluate the scheduling performance from $AV$ density and $DP$ density.
\end{enumerate}

The remainder of our manuscript is organized as follows. Sections \uppercase\expandafter{\romannumeral2} and \uppercase\expandafter{\romannumeral3} \ respectively review the related work and introduce preliminaries. Network model and design goals are presented in Section \uppercase\expandafter{\romannumeral4}. The implementation of SecLAVP is elaborated in Sections \uppercase\expandafter{\romannumeral5}. Security analysis and performance evaluation are provided in Sections \uppercase\expandafter{\romannumeral6} and \uppercase\expandafter{\romannumeral7}, respectively. At last, Section \uppercase\expandafter{\romannumeral8} concludes our work.
\section{Related Work}
\subsection{AVP Scheduling}
Uhlemann \cite{4} expounded the application prospect of valet parking by introducing Ford's remote-control repositioning and Volvo's pickup and dropoff service. Next, a distributed control method \cite{5} was utilized to coordinate parking spaces and driving paths in AVP. Timpner et al. \cite{6} proposed a V-Charge project to implement AVP and charging scheduling simultaneously, which was validated via parking statistics from Hamburg Airport and Braunschweig. Qin et al. \cite{7} proposed a path planning scheme for AVP, which respectively applied the hybrid A* algorithm and path coordination strategy to determine the global path and the transitional path.

The above schemes \cite{4,5,6,7} focus on SAVP, that is, passengers must drop off at the parking lot (semi-autonomous parking). Conversely, LAVP is fully autonomous parking because it allows passengers to be dropped off at a $DP$ (i.e. Drop-off/pick-up Point) near the destination, reducing the walking distance of passengers. LAVP has been concerned, but up to now, it is still in its infancy.

Given the parking price, fuel consumption, and distance from the parking lot, a cost-effective scheduling \cite{8} was proposed, which doesn't support parking reservations resulting in a low parking success rate. Therefore, based on the weighted value and heuristic method, Khalid et al. \cite{9} proposed a reservation scheme to minimize the walking distance of passengers and driving distance from the drop-off point to the parking lot. Our prior work \cite{10} proposed a reservation scheme to minimize passengers' travel time and $AV$ waiting time. Subsequently, from the perspective of electric vehicle charging reservation and parking waiting time, Zhang et al. \cite{11} jointly optimized energy consumption and travel services.

In summary, references \cite{4,5,6,7} emphasize the path planning in SAVP. References \cite{8,9,10,11} highlight parking reservations in LAVP. However, none of schemes \cite{4,5,6,7,8,9,10,11} concern communication security in AVP. Tampered or forged messages (such as parking requests and parking reservations) can affect the accuracy of path planning and parking space matching.
\subsection{AVP Security}
Huang et al. \cite{14} proposed a privacy protection reservation, preventing $AVs$ from reserving multiple parking spaces at the same time by zero-knowledge proof. However, the scheme can't resist the conspiracy between parking lot terminals and parking service providers. Additionally, this scheme ignores that message may be tampered with or forged (i.e. confidentiality and unforgeability).

To defend against attackers stealing or remotely controlling $AVs$, Ni et al. \cite{15} proposed a two-factor AKA based on passwords and smartphones. Furthermore, BBS+ signature and Cukoo filter were applied to anonymous authentication and store pickup codes. However, with the increment of $AV$ number, communication costs between local server and parking service provider increases sharply. Additionally, confidentiality and unforgeability are ignored.

Pokhrel et al. \cite{16} utilized differential privacy and zero-knowledge proof to protect users' location and identity privacy. To further improve the parking reservation rate, Markov decision-based reinforcement learning was applied to optimize parking space matching. Nonetheless, this work doesn't authenticate the user's identity in advance causing illegal access. Besides, multiple parking spaces are simultaneously reserved resulting in poor QoS.

To prevent $AV$ theft and location leakage, Fang et al. \cite{17} proposed a privacy-preserving AVP based on blockchain (named B-park), which was upgrade of Pointcheval-Sander (PS) group signature. B-park was deployed in Etherums to realize anonymity and traceability, whereas this scheme can't resist sybil attacks and man-in-middle attacks.

Xu et al. \cite{18} proposed a lattice-based ring signature to resist privacy leakage under quantum attacks. This solution not only protects identity privacy, but also satisfies unforgeability, unlinkability, and quantum attacks resistance.

In short, the above schemes focus on the privacy protection of SAVP, ignoring confidentiality and unforgeability, as well as excessive overheads. Additionally, the above schemes are concentrated on parking request verification and parking check-in, which aren't applicable LAVP because they ignore identity legitimacy and communication security in the phase of picking up as well as dropping off passengers.
\subsection{AKA Protocol}
Chien and Wu \cite{19} proposed a passwords-based AKA protocol for secure authentication between users and servers. This protocol enables efficient addition of users, but doesn't consider user revocation. With the prosperity of end-to-end communication, Byun et al. \cite{20} proposed a passwords-based AKA protocol for client-to-client. Subsequently, Turkanovi{\'c} et al. \cite{21} proposed an AKA for heterogeneous wireless networks, which doesn't defend against smart card loss and sensor node spoofing attacks. Seeking to address the above challenges, Chang and Le \cite{22} proposed a provably secure authentication scheme. Srinivas et al. \cite{23} proposed a two-factor AKA based on smart card and passwords for healthcare. However, communication overheads are large.

The above schemes focus on single factor AKA \cite{19,20} and two-factor AKA \cite{21,22,23}, facing issues such as smart card loss, passwords guessing, and biometrics leakage. Subsequently, three-factor (smart card, passwords, and biometrics) AKA has attracted widespread attention.

To ensure reliable and secure charging from the smart grid, Wazid et al. \cite{24} proposed an authentication scheme for secure communication between electric vehicles and charging piles. Nonetheless, this solution doesn't support the revocation of charging piles captured by attackers. To prevent biometrics leakage, Zhang et al. \cite{25} proposed a dynamic authentication based on biohash and hash functions to verify biometrics. Nonetheless, the dynamic verification table stored locally by the server has a large overhead. Srinivas et al. \cite{26} proposed a biohashing-based AKA scheme integration of passwords and biometrics, which effectively resists impersonate attacks. For agriculture application, Ali et al. \cite{27} proposed a remote AKA to prevent monitoring data (such as temperature and humidity) from being illegally accessed and tampered with. However, solutions \cite{26,27} can't resist offline dictionary attacks and internal attacks. Li et al. \cite{28} proposed a robust three-factor authentication protocol based on a lightweight hash function for the Industrial Internet of Things, in which visitors can access sensors after being authenticated by gateways. Nevertheless, this scheme can't resist offline dictionary attacks. Jiang et al. \cite{29} combined biometrics, passwords, and smart card to achieve authentication between a cloud center, $AVs$, and users. This scheme prevents attackers from tampering, forging, and eavesdropping on data, but it can't resist offline dictionary attacks and has large communication overheads. Sureshkumar et al. \cite{42} proposed a three-factor AKA protocol for vehicle-to-grid to realize secure communication between electric vehicles and the smart grid, but this scheme can't update biometrics. Based on the chaotic-maps discrete logarithm problem, Srinivas et al. \cite{30} proposed a three-factor AKA, which was proven in ROR. Subsequently, based on the chaotic-maps computational Diffie-Hellman problem, Qiu et al. \cite{31} introduced fuzzy-verifier and honeywords to achieve three-factor security.

In conclusion, the above schemes \cite{26,27,28,29} still face smart card loss, passwords guessing, and internal attacks. Additionally, the solution \cite{25} lacks formal security verification and has a large overhead. Further to say, references \cite{24,28,29,42} lack formal security proof and formal security verification. Due to its advantages of unpredictability and uncertainty, chaotic maps are introduced into AKA \cite{30,31}, but the computational efficiency is still not high. Table \uppercase\expandafter{\romannumeral1} compares state-of-the-art AKA protocols.

Therefore, existing three-factor AKA protocols must balance security and availability. In other words, we should not only address potential security threats, provide formal security proof and formal security verification, but also minimize overheads.
\begin{table*}\scriptsize
 \centering
 \caption{State-of-the-art AKA protocols comparison}
 \begin{tabular}{|c|l|l|p{3.65cm}|p{4cm}|p{4.5cm}|}
\hline
Category &\multicolumn{1}{c|}{Scheme} &\multicolumn{1}{c|}{Year} &\multicolumn{1}{c|}{Primitives} &\multicolumn{1}{c|}{Advantages} &\multicolumn{1}{c|}{Limitations}\\
\hline
\multirow{2}{*}{Single factor} & Chien and Wu\cite{19} &2009 &\ding{172} Hash function \ \ding{173} XOR operation \newline\ding{174} Elliptic curve cryptography ($ECC$) & Dynamic user addition. &Dynamic user revocation isn't supported. \\
\cline{2-6}
& Byun et al. \cite{20} &2007 &\ding{172} Hash function \newline\ding{173} Symmetric encryption & User authentication for two different domains. &Users cannot register and authenticate on multiple different servers. \\
\hline
\multirow{3}{*}{Two-factor} & Turkanovi{\'c} \cite{21} &2014 &\ding{172} Hash function \ \newline\ding{173} XOR operation & Lightweight. &Can't resist smart card loss and sensor node spoofing attacks. \\
\cline{2-6}
& Chang and Le \cite{22} &2016 &\ding{172} Hash function \ \newline\ding{173} XOR operation & Resist smart card loss and sensor node spoofing attacks. &No communication overhead comparison. \\
\cline{2-6}
& Srinivas et al. \cite{22} &2020 &\ding{172} Hash function \ \ding{173}XOR operation & Resist privileged-insider attacks. & High communication overheads. \\
\hline
\multirow{8}{*}{Three-factor} & Wazid et al. \cite{24} &2017 &\ding{172} Hash function \newline\ding{173} XOR operation \newline\ding{174} $ECC$ & Secure communication between users and smart meters. &Malicious charging piles revocation isn't supported. No formal security proof and verification. \\
\cline{2-6}
& Zhang et al. \cite{25} &2018 &\ding{172} Hash function \ding{173} Biohash function  \newline\ding{174} XOR operation & Users' biometrics privacy protection. &Dynamic verification table brings in high overheads. No formal security verification. \\
\cline{2-6}
& Srinivas et al. \cite{26} &2017 &\ding{172} Hash function \ding{173} Biohash function  \newline\ding{174} XOR operation & Resist smart card loss and protect biometrics privacy. & No formal security proof. \\
\cline{2-6}
& Ali et al. \cite{27} &2018 &\ding{172} Hash function \ding{173} Biohash function  \newline\ding{174} Fuzzy extractor \ding{175} XOR operation & Support anonymity and resist reply attacks. & Can't resist offline dictionary attacks and can't support forward security. \\
\cline{2-6}
& Li et al. \cite{28} &2018 &\ding{172} Hash function \newline\ding{173} XOR operation  \newline\ding{174} $ECC$ & Support anonymity and passwords update. Resist reply attacks. & Smart card revocation and biometrics update aren't supported. No formal security proof and verification. \\
\cline{2-6}
& Jiang et al. \cite{29} &2020 &\ding{172} Hash function \ \ding{173} XOR operation  \newline\ding{174} Biometric extraction & Secure communication between users, cloud, and autonomous vehicle. & Can't resist offline dictionary attacks. No formal security proof and verification. \\
\cline{2-6}
& Sureshkumar et al. \cite{42} &2022 &\ding{172} Hash function \newline \ding{173} XOR operation & Support passwords update and recovery. & Can't support biometrics update, formal security proof, and formal security verification. \\
\cline{2-6}
& Srinivas et al. \cite{30} &2020 &\ding{172} Hash function \ \ding{173} Fuzzy extractor  \newline\ding{174} XOR operation & Support passwords and biometrics update, as well as smart card revocation. & Can't support forward security and low computational efficiency. \\
\cline{2-6}
& Qiu et al. \cite{31} &2022 &\ding{172} Hash function \ding{173} XOR operation  \newline\ding{174} Fuzzy extractor \ding{175} Honeywords & Secure communication between users and server. & Low computational efficiency. \\
\hline
\end{tabular}
\end{table*}

\section{Preliminaries}
\subsection{Bilinear Pairing}
Note that $G_1$ and $G_2$  are multiplicative cyclic groups with the prime order $p$ . The generator of $G_1$ is $g$. Bilinear pairing \cite{34} $e:{G_1} \times {G_1} \to {G_2}$ satisfies the following properties.

1)\ \textit{Bilinearity}: $e\left(V^{\lambda}, W^{\zeta}\right)=e\left(V^{\zeta}, W^{\lambda}\right)=e(V, W)^{\lambda \zeta}$ for any $V, W \in G_{1}$ as well as $\lambda, \zeta \in Z_{p}^{*}$.

2)\ \textit{Computability}: for any $V, W \in G_{1}$, bilinear pairing $e(V, W)$ is capable of being computed effectively.

3)\ \textit{Non-degeneracy}: $e(V, W) \neq 1$ for all $V, W \in G_{1}$.
\subsection{Biometric Extraction}
In 2004, Dodis et al. \cite{35} first proposed the fuzzy extractor, which can not only improve the matching efficiency of biometrics, but also prevent the exposure of biometrics. The fuzzy extractor consists of the following two algorithms.

1)\ $(\alpha, \beta) \leftarrow\mathrm {\emph{KPGen}}(Bio)$: Given the biometrics $Bio$ and the public parameter generation algorithm {\emph{KPGen}}, output the biometric key $\alpha$ and the public parameter $\beta$.

2)\ $\alpha\leftarrow$ $\operatorname{\emph{KeyRet}}\left(\right.Bio\left.^\prime,\beta\right)$: The biometrics $Bio^\prime$  and the public parameter $\beta$ are input, then the biometric retrieval algorithm {\emph{KeyRet}} generates the biometric key $\alpha$ only if the equation   ${dis}\left(\right.Bio^\prime,Bio) \leq \varepsilon$ is satisfied.   $dis$ and $\varepsilon$ represent statistical distance and specified distance, respectively.

\section{Network Model and Design Goals}
\subsection{System Model}
Fig. 1 presents the system model and  security threats, entities involving $DPs$, $AVs$, passengers, and parking lots. The above entities are defined in detail below.

\begin{figure}[htbp]
\setlength{\abovecaptionskip}{-0.1cm}
\setlength{\belowcaptionskip}{-0cm}
\centering
\includegraphics[scale=0.33]{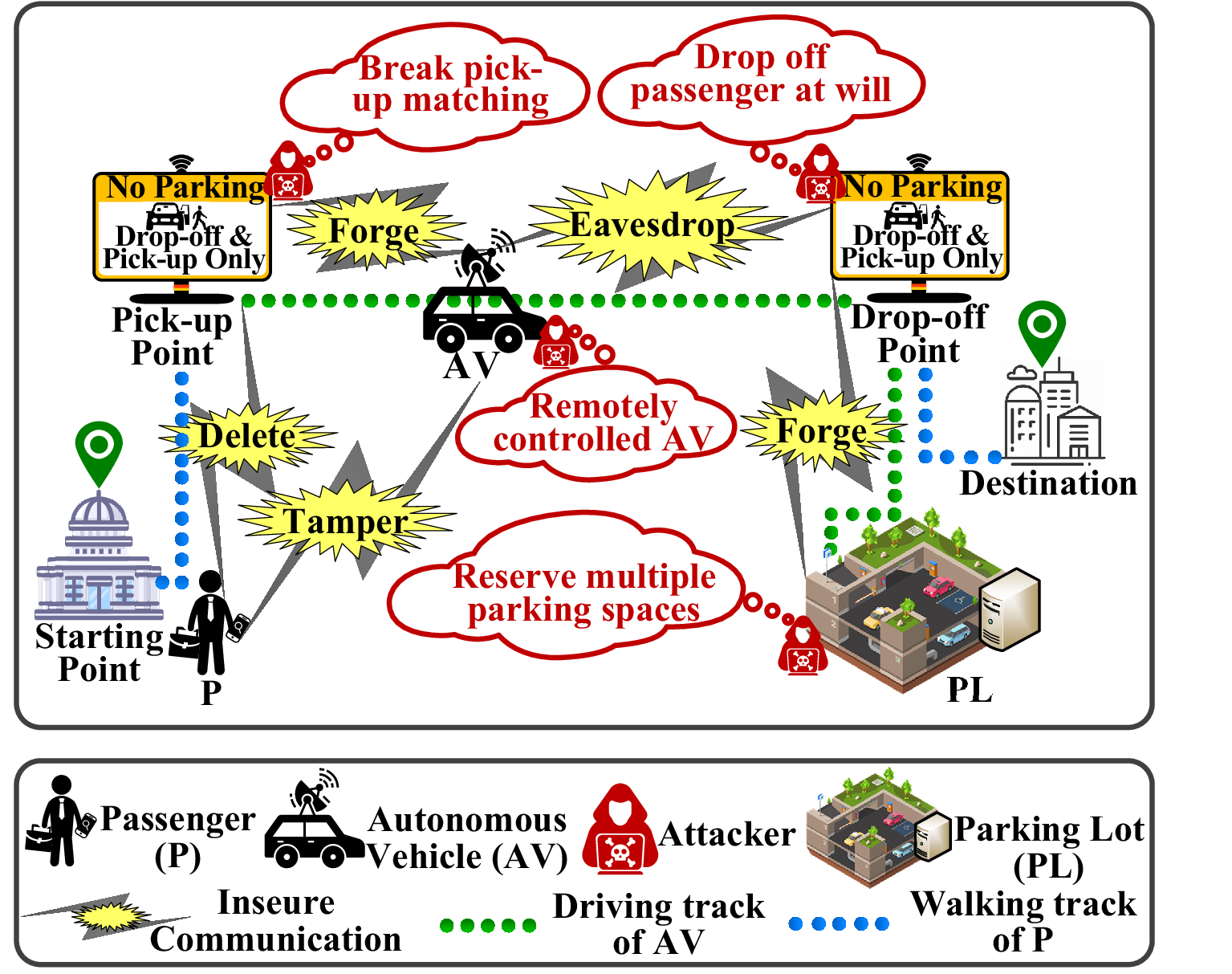}
\caption{System model and security threats}
\end{figure}

1)\textbf{\ Drop-off/pick-up points ($DPs$)}. $DPs$ only allow autonomous vehicles ($AVs$) to drop off or pick up passengers, and do not allow $AVs$ to stay for a long time. $DPs$ are responsible for passenger registration and $AV$ registration. More than that, $DPs$ collect and process parking reservations, which then determine the best drop-off point, suitable parking lot, and optimal $AV$.

2)\textbf{\ Autonomous vehicles ($AVs$)}. $AVs$ are registered with $DPs$. When the $AV$ receives the matching results (i.e. pick-up point, the best drop-off point, optimal parking lot, and matching passenger), the $AV$ first drives to the pick-up point to pick up the passenger. Then the $AV$ delivers the passenger to the drop-off point and finally travels to the parking lot for autonomous valet parking.

3)\textbf{\ Passengers}. Passengers get smart cards after registering at $DPs$. Passengers who need to ride send ride requests to adjacent $DPs$, and then walk to adjacent $DPs$ to wait for $AV$ pick-up.

4)\textbf{\ Parking lots}. Equipped with cameras, sensors, and communication facilities, parking lots ($PLs$) supervise and manage the status of parking spaces, and then feed back parking spaces' real-time status to $DPs$.

\subsection{Security Model}
The Real-Or-Random (ROR) model \cite{32} is utilized to prove the session key security of our proposed SecLAVP. The ROR model has been widely used for formal security proof in existing AKA protocols \cite{22,24,25,30,31}. Primitives in the ROR model are defined as follows.

\textit{Participants}. $\Pi$ represents an instance (i.e. oracles). $\Pi_{{P_i}}^i$, $\Pi_{{DP_i}}^j$, and $\Pi_{{AV_i}}^k$ denote $i-th$, $j-th$, $k-th$ instances of $P_i$, $DP_i$, and $AV_i$, respectively.

\textit{Accepted state}. If an instance $\Pi$ receives the expected message, it enters accepted state. Moreover, the identity ($SID$) of the current session consists of all sent and received messages.

\textit{Partnering}. Two instances ${\Pi ^i}$ and ${\Pi ^k}$ satisfy the following three conditions, then ${\Pi ^i}$ and ${\Pi ^k}$ are partners.

1)\ ${\Pi ^i}$ and ${\Pi ^k}$ complete mutual authentication.

2)\ ${\Pi ^i}$ and ${\Pi ^k}$ are accepted state.

3)\ ${\Pi ^i}$ and ${\Pi ^k}$ are each other's partners.

\textit{Freshness}. Supposing that an adversary $\mathcal{A}$ can't obtain the session key between $P_i$ and $AV_i$ through the Reveal query defined in the reference \cite{22}, then $\Pi _{{P_i}}^i$ or $\Pi _{A{V_i}}^k$ is freshness.

\textit{Queries}. The adversary $\mathcal{A}$ implements various attacks via the following queries.

\begin{itemize}
 \item $Execute(\Pi _{{P_i}}^i,\Pi _{D{P_i}}^j,\Pi _{A{V_i}}^k)$: $\mathcal{A}$ eavesdrops on messages between $P_i$, $DP_i$, and $AV_i$ by $Execute(\Pi _{{P_i}}^i,\Pi _{D{P_i}}^j,\Pi _{A{V_i}}^k)$ query, which simulates passive attacks.
 \item ${\mathop{\rm Re}\nolimits} veal(\Pi )$: $\mathcal{A}$ obtains the current session key of via the ${\mathop{\rm Re}\nolimits} veal(\Pi )$ query.
 \item $Send(\Pi ,Mes)$: $\mathcal{A}$ sends a forged message to $\Pi$ to get a response by the $Send(\Pi ,Mes)$ query, which simulates active attacks.
 \item $CorruptSC(\Pi _{{P_i}}^i)$: $\mathcal{A}$ obtains secret parameters stored in the stolen smart card for authenticating the passenger through $CorruptSC(\Pi _{{P_i}}^i)$ query, which simulates smart card loss attacks.
 \item $CorruptAV(\Pi _{{AV_i}}^k)$: The long-term secret key is obtained by $\mathcal{A}$ through the $CorruptAV(\Pi _{{AV_i}}^k)$ query, which is used to prove perfect forward security.
 \item $Test(\Pi)$: Based on the indistinguishability of the ROR model, the $Test(\Pi)$ query is used to prove the semantic security of the session key. A bit $\ell$ is randomly selected during experiment initialization to determine the output of $Test(\Pi)$. $\mathcal{A}$ executes $Test(\Pi)$ query. Only if the session has been established and is freshness, the session key is output when $\ell=1$, or else a random string is output. Otherwise, the $Test(\Pi)$ query outputs $\bot$.
\end{itemize}

Under the ROR model, $\mathcal{A}$ must be able to distinguish between the real session key and the random key. $\mathcal{A}$ conducts a $Test()$ query regarding $\Pi _{{P_i}}^i$ or $\Pi _{A{V_i}}^k$, and the output of the $Test()$ query must be consistent with $\ell$. If the guess bit $\ell^{'}$ output by $\mathcal{A}$ is equal to $\ell$, $\mathcal{A}$ wins the game (The event is represented by ${E_{Suc}}(\mathcal{A})$).

\textbf{Definition 1. AKA security (i.e. \textit{semantic security of the session key}).} Let
\begin{equation}
Adv_{\mathcal{A}}^{AKA}(pt) = |2\Pr [{E_{Suc}}({\mathcal{A}})] - 1|
\nonumber
\end{equation}
where $Adv_{\mathcal{A}}^{AKA}$ is the advantage of the adversary $\mathcal{A}$ breaking the semantic security of the session key (i.e. AKA security). Furthermore, $\Pr [{E_{Suc}}({\mathcal{A}})]$ is the probability of the event ${E_{Suc}}(\mathcal{A})$. In the ROR model, the protocol is AKA security if $Adv_{\mathcal{A}}^{AKA}$ is negligible in polynomial time $pt$.

\textbf{Definition 2. Elliptic curve decisional Diffie-Hellman ($ECDDH$) problem.} $F_p$ and $P$ are respectively the finite field of prime order $p$ and the generator of the cyclic group. Elliptic cyclic group ${E_{a,b}}:{y^2} = {x^3} + ax + b$ satisfies $4{a^3} + 27{b^2} \ne 0\bmod p$. Given $uP$, $vP$, and $wP$, we can determine whether the equation $uvP=wP$ holds, where $u,v,w \in Z_p^*$.
\subsection{Adversary Model}
The Dolev-Yao ($DY$) threat model \cite{36} is applied to SecLAVP, in which communication parties transmit information through public channel. The adversary can control public channel, that is to say, he/she can tamper, forge, inject or delete information, as well as intercept and eavesdrop on information. Besides, the adversary obtains the parameters stored in the smart card through side channel or energy analysis when the smart card is lost. Now that Canetti and Krawczyk's adversary model ($CK$-adversary model) \cite{37} is regarded as a standard model for simulating AKA protocols, we introduce it into SecLAVP. Compared with the $DY$ threat model, the adversary in the $CK$-adversary model can capture the previous session keys. Note that selfish $AVs$ will meanwhile reserve multiple parking spaces for maximizing parking reservation rate.

\subsection{Design Goals}
Following our proposed system model, security model, and adversary model, the following design goals $DG_{i}$ must be met.

1)\ \textbf{\bm{$DG_{1}$} Anonymity}: $DPs$ can't detect passengers' identities and can't infer two sessions sent by the same user.

2)\ \textbf{\bm{$DG_{2}$} Mutual Authentication}: Communication entities must authenticate each other to ensure secure communication.

3)\ \textbf{\bm{$DG_{3}$} AKA Security}: Passengers and $DPs$ negotiate to generate the session key, so do $AVs$ and $DPs$.

4)\ \textbf{\bm{$DG_{4}$} Forward Security}: Although attackers obtain current session key, they can't calculate previous session key.

5)\ \textbf{\bm{$DG_{5}$} Update}: Support passwords update, biometrics update, and smart card revocation.

6)\ \textbf{\bm{$DG_{6}$} No Passwords Verified Table}: $DPs$ and $AVs$ don't store sensitive information related to passwords.

7)\ \textbf{\bm{$DG_{7}$} Passwords Security}: $DPs$ can't infer passwords.

8)\ \textbf{\bm{$DG_{8}$} Anti Smart Card Loss}: Passenger's secret parameters won't be disclosed when the smart card is lost.

9)\ \textbf{\bm{$DG_{9}$} Internal Attacks Resistance}: Passengers won't disclose sensitive information to $DPs$ in the registration.

10)\ \textbf{\bm{$DG_{10}$} Replay Attacks Resistance}: Attackers can't repeatedly send ride requests to deceive $DPs$.

11)\ \textbf{\bm{$DG_{11}$} Offline Dictionary Guessing Attacks Resistance}: Attackers can't guess the passenger's identity and passwords by a custom dictionary.

12)\ \textbf{\bm{$DG_{12}$} Multiple-reservation Resistance}: Only one parking space can be successfully reserved at a time.

13)\ \textbf{\bm{$DG_{13}$} Efficiency}: On the premise of ensuring security, we should minimize overheads and optimize scheduling performance.

14)\ \textbf{\bm{$DG_{14}$} Formal Security Proof}: Formal security proof is indispensable to strictly ensure the security of the protocol.

15)\ \textbf{\bm{$DG_{15}$} Formal Security Verification}: The official protocol verification software shall be used for verifying security.

\section{The Proposed Scheme}

\subsection{Overview}

This section expounds on the implementation process of SecLAVP, mainly including Setup, Pick up, Drop off \& Parking, as well as Update \& Revocation, as shown in Fig. 2. Key symbols are shown in Table \uppercase\expandafter{\romannumeral2}.
\begin{figure}[htbp]
\setlength{\abovecaptionskip}{-0.1cm}
\setlength{\belowcaptionskip}{-0cm}
\centering
\includegraphics[scale=0.39]{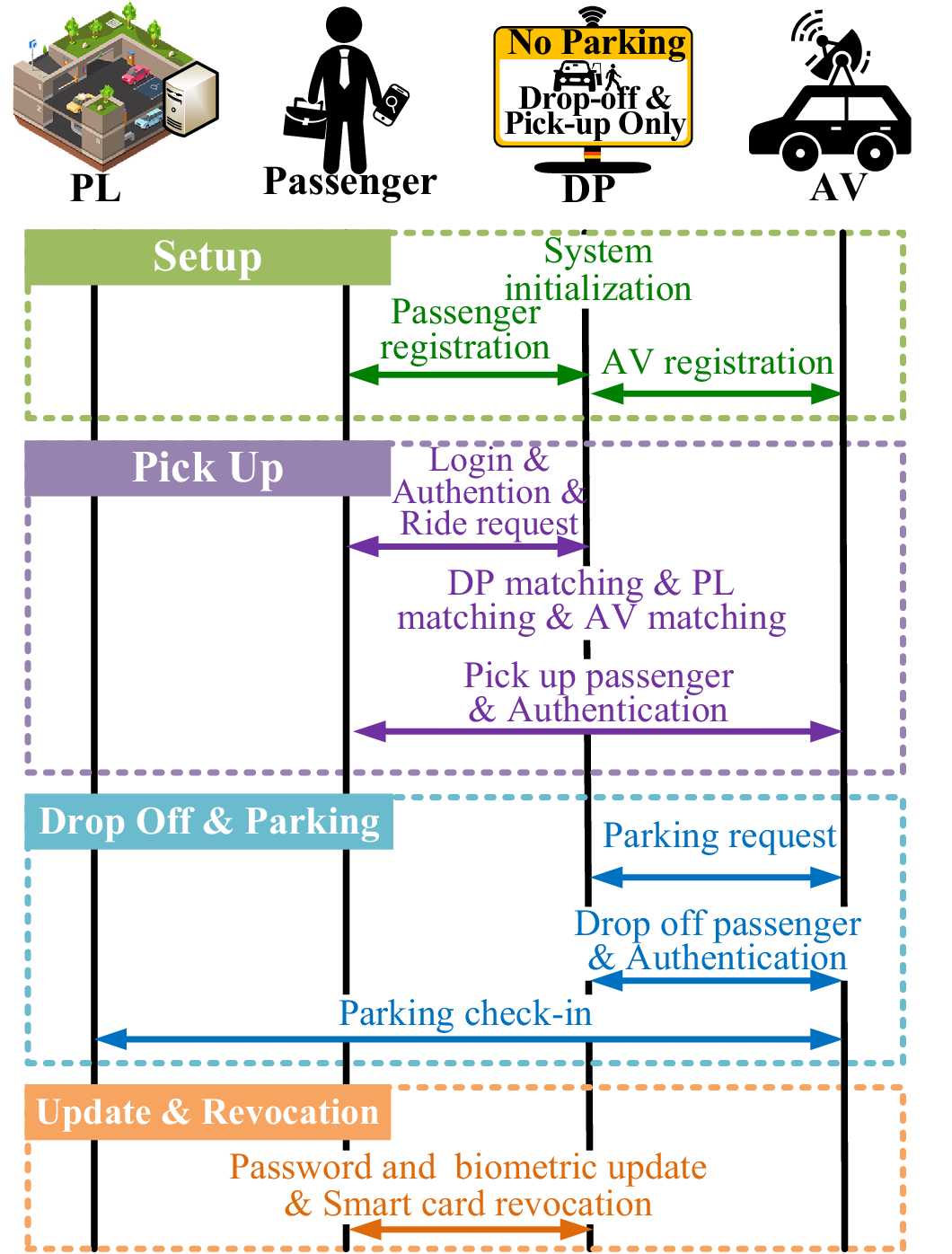}
\caption{The implementation process diagram of SecLAVP}
\end{figure}

1)\ \textbf{Setup}: {\large\ding{172}} The parameters required by the system are generated in system initialization. {\large\ding{173}} Passengers register at the adjacent $DP$. {\large\ding{174}} $AVs$ register at the adjacent $DP$.

2)\ \textbf{Pick Up}: {\large\ding{172}} Passengers who have completed registration and need to take a ride must login and authenticate. After successful authentication, a session key is generated and a ride request is sent to the adjacent $DP$. {\large\ding{173}} $DP$ matches passengers with the best drop-off point, optimal parking lot and suitable $AV$ after receiving the ride request. {\large\ding{174}} The matched $AV$ drives to the pick-up point and authenticates with the passenger to ensure picking up a matched passenger.

3)\ \textbf{Drop Off \& Parking}: {\large\ding{172}} The $AV$ sends parking request and parking reservation to $DPs$ of route. {\large\ding{173}} $DPs$ of route send the best drop-off point and optimal parking lot to the $AV$. The $AV$ drives to the best drop-off point and verifies the best drop-off point. After the verification is passed, the $AV$ drops off the passenger. {\large\ding{174}} The $AV$ drives to matching parking lot for parking check-in.

4)\ \textbf{ Update \& Revocation}: {\large\ding{172}} Passengers update passwords and biometrics locally. {\large\ding{173}} Revoke smart card when the passenger loses smart card.
\begin{table}[htbp]\scriptsize%footnotesize
 \setlength{\abovecaptionskip}{-0.05cm}
 \setlength{\abovecaptionskip}{-0.05cm}
 \centering
 \caption{Key symbols}
 \label{table_time}
 \begin{spacing}{1.12}
 \begin{tabular}{|p{3.2cm}|p{4.8cm}|}
  \hline
  Symbol & Description\\
  \hline
  $P_i$ & the passenger $P_i$\\
  \hline
  $ID_{P_i};$ $PW_{P_i};$ $Bio_{P_i}$ & ${P_i}'s$ identity, passwords, and biometrics\\
  \hline
  $DP_i$ & the neighboring pick-up point\\
  \hline
  $G_3$; $P$ & additive cyclic groups; the generator\\
  \hline
  $ID_{DP_k}$; $ID_{PL_i}$; $ID_{AV_i}$ & the identities of the best drop-off point, the optimal $PL$  as well as the optimum $AV$\\
  \hline
  $ADSK$ & the session key between $AV_i$ and $DP_i$\\
  \hline
  $PDSK$ & the session key between $P_i$ and $DP_i$\\
  \hline
  $PASK$ & the session key between $P_i$ and $AV_i$\\
  \hline
  $h()$; $||$; $\oplus$ & hash function; concatenation; XOR operation\\
  \hline
  $Des$ & ${P_i}'s$ destination\\
  \hline
  $T_{Parking};$ $t_{Parking}$ & parking moment; parking duration\\
  \hline
  $L$ & the list counting the number of inputting wrong passwords\\
  \hline
 \end{tabular}
 \end{spacing}
\end{table}
\subsection{Setup}
\noindent {1)\textbf{\ System Initialization}}

First, $DPs$ select additive cyclic groups $G_3$ with the prime order $p$. Second, $DPs$ randomly select $x_{DP}$ as the private key to calculate the public key $X_{DP}=x_{DP} P$, where $P$ is the generator of $G_3$. Supposing that $ID_{DP}$ is the identity of $DP$  and $\left\{x_{D P}, I D_{D P}\right\}$ is stored locally by $DPs$. Finally, $DPs$ assign a unique identity $ID_{AV}$ to the autonomous vehicle $AV$. The $AV$ and $DPs$ store $ID_{AV}$ locally.

\noindent {2)\textbf{\ Passenger Registration}}

Fig. 3 shows the detailed process of passenger registration (hereinafter referred to as $PR$) in SecLAVP.
\begin{figure}[htbp]
\setlength{\abovecaptionskip}{-0.1cm}
\setlength{\belowcaptionskip}{-0cm}
\centering
\includegraphics[scale=0.42]{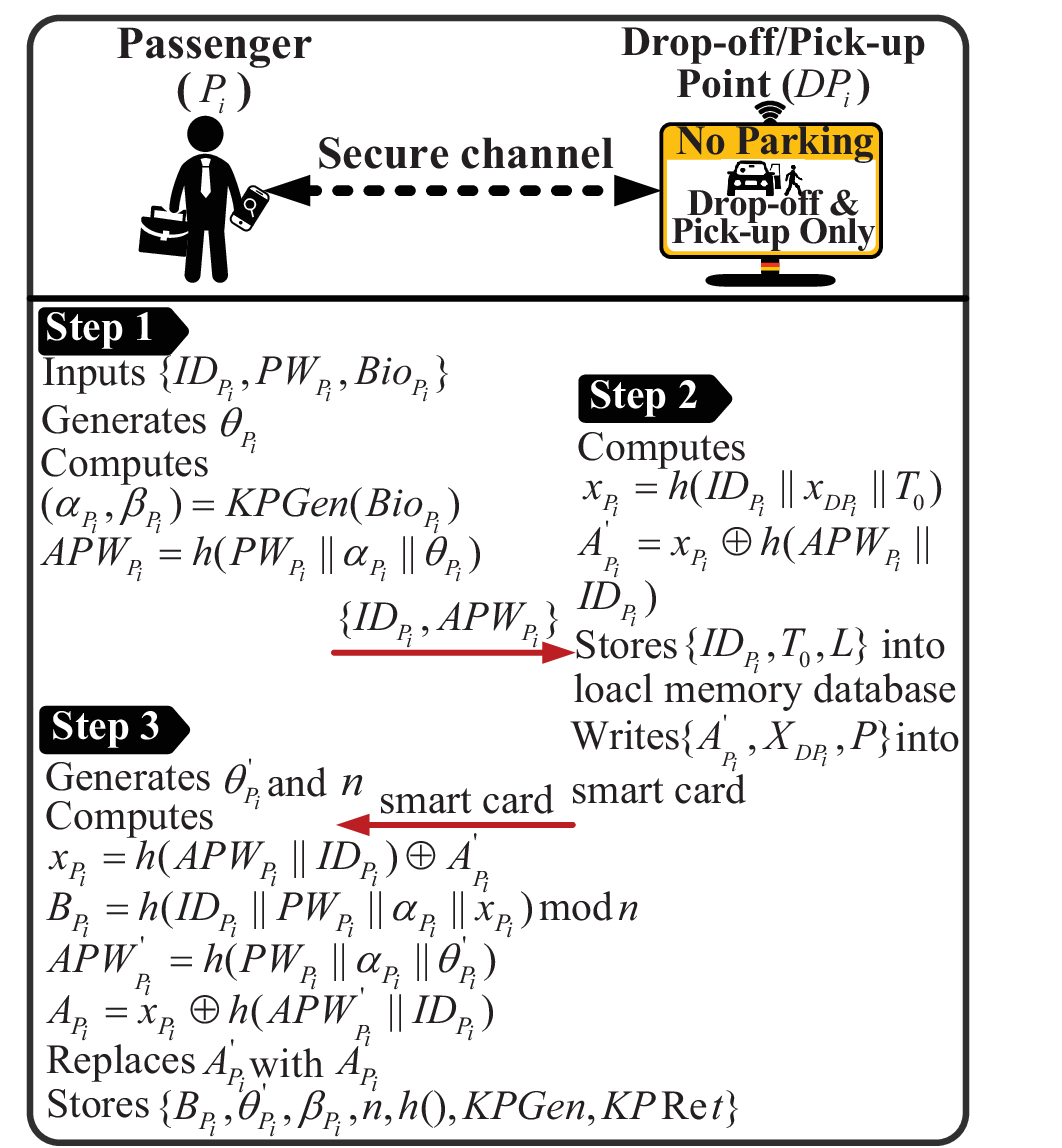}
\caption{Passenger registration}
\end{figure}

$\bm{Step \ PR_1}$: The passenger $P_{i}$ chooses $ID_{P_{i}}$ as the identity and sets custom string $PW_{P_{i}}$ as own passwords. Next, $P_{i}$ calculates the biometric key  $\alpha_{P_{i}}$ and public parameter $\beta_{P_{i}}$ via biometric extraction (see the Section \uppercase\expandafter{\romannumeral3}. B for details), namely $\left(\alpha_{P_{i}}, \beta_{P_{i}}\right)=KPGen\left(Bio_{P_{i}}\right)$. Eventually, $P_{i}$ chooses a random number $\theta_{P_{i}}$ to calculate $APW_{P_{i}}=h\left(P W_{P_{i}}\left\|\alpha_{P_{i}}\right\| \theta_{P_{i}}\right)$ and forwards $\left\{ID_{P_{i}}, APW_{P_{i}}\right\}$ to the adjacent $DP$ (hereinafter defined as $DP_{i}$).

$\bm{Step\ PR_2}$: $DP_{i}$ checks whether $ID_{P_i}$ exists in the local database after receiving $\left\{ID_{P_{i}}, APW_{P_{i}}\right\}$. $P_{i}$ must reselect the identity if $ID_{P_i}$ already exists, or else $DP_{i}$ calculates $x_{P_{i}}=h\left(ID_{P_{i}}\left\|x_{DP_{i}}\right\|T_{0}\right)$ and $A_{P_{i}}^{\prime}=x_{P_{i}}\oplus h\left(APW_{P_{i}}\|ID_{P_{i}}\right)$, where $T_0$ is the timestamp. $DP_i$ then creates a list $L$ to count the number of inputting wrong passwords. Note that the initial value and the threshold value of $L$ are 0 and 5. Note that the smart card will be frozen for 30 minutes as long as $L$ is greater than 5. After $DP_{i}$ stores $\left\{ID_{P_{i}},T_{0}\right\}$ and $L$ locally, $\left\{A_{P_{i}}^\prime,X_{D P_{i}},P\right\}$ is written into the smart card, which is sent to $P_i$.

$\bm{Step\ PR_3}$: $P_{i}$ calculates $x_{P_{i}}=h\left(APW_{P_{i}}\|ID_{P_{i}}\right)\oplus A_{\mathrm{P_{i}}}^\prime$, $B_{P_{i}}=h\left(ID_{P_{i}}\left\|PW_{P_{i}}\right\|\alpha_{P_{i}}\|x_{P_{i}}\right)\bmod n$, $APW^\prime_{P_{i}}=h\left(P W_{P_{i}}\left\|\alpha_{P_{i}}\right\| \theta^\prime_{P_{i}}\right)$, and $A_{P_{i}}=x_{P_{i}}\oplus h\left(APW^\prime_{P_{i}}\|ID_{P_{i}}\right)$. $\theta^{'}_{P_{i}}$ and $n$ are random numbers. $\left\{A_{p_{i}},B_{P_{i}},\theta_{P_{i}}^{\prime},\beta_{P_{i}},X_{DP_{i}},P,n,h(),KPGen,KPRet\right\}$ is stored in the smart card $SC$.

\noindent {3)\textbf{\ \bm{$AV$} Registration}}

$AV$ sends the identity $ID_{AV}$ and a registration request $Req_{AV}$ to the $DP$, which calculates the private key $x_{A V}=h\left(I D_{A V} \| x_{D P}\right)$ and sends $x_{AV}$ to $AV$ by a secure channel.

\subsection{Pick Up}
\noindent {1)\textbf{\  Login}}
\begin{figure*}[htb]
\setlength{\abovecaptionskip}{-0.1cm}
\setlength{\belowcaptionskip}{-0cm}
\centering
\includegraphics[scale=0.4]{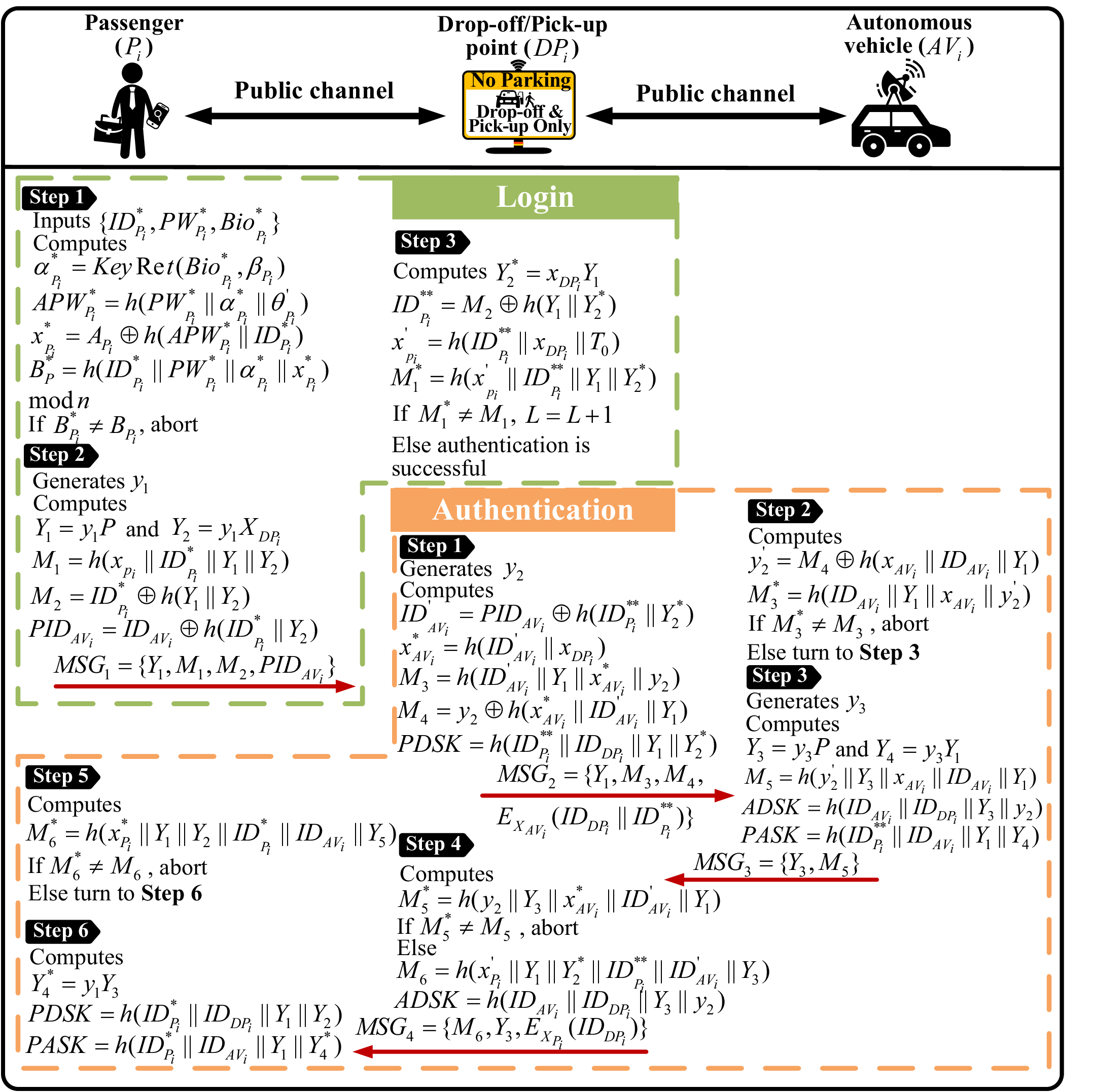}
\caption{Login and authentication}
\end{figure*}

Fig. 4 presents the detailed step of login (hereinafter referred to as $Lg$) in the green dotted box.

$\bm{Step\ Lg_1}$: The smart card $SC$ calculates the biometric key $\alpha_{P_{i}}^{*}={KeyRet}\left(Bio^{*}_{{P}_{i}},\beta_{P_{i}}\right)$ after $P_i$ inputs the identity $ID^*_{P_{i}}$, passwords $PW^*_{P_i}$, and biometrics $Bio^*_{P_i}$. Afterwards, $SC$ calculates $APW^{*}_{P_{i}}=h\left(PW^{*}_{P_{i}}\left\|\alpha^{*}_{P_{i}}\right\| \theta^\prime_{P_{i}}\right)$, $x^{*}_{P_{i}}= A_{P_{i}}\oplus h\left(APW^{*}_{P_{i}}\|ID_{P_{i}}\right)$ as well as $B^{*}_{P_{i}}=h(ID^{*}_{P_{i}}||PW^{*}_{P_{i}}||\alpha^{*}_{P_{i}}||x^{*}_{P_{i}})mod \ n$ and verifies whether the equation $B^*_{P_i}=B_{P_i}$ holds. Once the above equation is established, turn to $\bm{Step\ Lg_2}$ and $DP_i$ sends the matched $AV's$ identity (i.e. $ID_{AV_i}$) to $P_i$, otherwise terminate login.

$\bm{Step\ Lg_2}$: $SC$ selects a random number $y_1$ to calculate $Y_1=y_{1}P$, $Y_2=y_{1}X_{DP_i}$, $M_{1}=h(x_{p_{i}}||ID_{P_{i}}^{*}||Y_{1}||Y_{2})$ $M_{2}=ID_{P_{i}}^{*} \oplus h(Y_{1}||Y_{2})$, and $PID_{AV_{i}}=ID_{AV_{i}}\oplus h(ID_{P_{i}}^{*}||Y_{2})$, where $ID_{AV_i}$ is obtained by $SC$ inquiring $P_i$. $P_i$ generates the message $MSG_1$ including $Y_1$, $M_1$, $M_2$, and $PID_{AV_i}$, then sends $MSG_1$ to $DP_i$.

$\bm{Step\ Lg_3}$: $DP_i$ calculates $Y^*_2=x_{DP_i}Y_{1}$, $ID_{P_{i}}^{**}=M_{2}\oplus h\left(Y_{1} \|Y_{2}^{*}\right)$, and $x_{p_{i}}^{\prime}=h\left(ID_{P_{i}}^{**}\left\|x_{DP_{i}}\right\|T_{0}\right)$ after receiving $MSG_1$. $DP_i$ then calculates $M^*_{1}=h(x^\prime_{p_{i}}||ID^{**}_{P_{i}}||Y_{1}||Y^*_{2})$. Finally, $DP_i$ checks whether the equation $M^*_1=M_1$ holds. Provided that the above equation set up, login is successful, or else login is terminated and $L$ is updated as $L=L+1$.

\noindent {2)\textbf{\ Authentication}}

Fig. 4 presents the detailed step of login (hereinafter referred to as $Au$) in the orange dotted box.

$\bm{Step\ Au_1}$: $AV_i$ sends $ACK$ to $DP_i$ after arriving at $DP_i$. $DP_i$ randomly selects $y_2$ to calculate $ID^{\prime}_{AV_{i}}=PID_{AV_{i}}\oplus h\left(ID^{**}_{P_{i}}\|Y^{*}_{2}\right)$ and $x^*_{AV_i}=h\left(ID^\prime_{AV_{i}}\|x_{DP_i}\right)$. Next, $DP_i$ calculates $M_{3}=h\left(I D_{A V_{i}}^{\prime}\left\|Y_{1}\right\| x_{A V_{i}}^{*} \| y_{2}\right)$ $M_{4}=y_{2} \oplus h(x_{AV_{i}}^{*}||ID_{A V_{i}}^{\prime}||Y_{1})$, and $PDSK=h\left(I D_{P_{i}}^{* *}\left\|I D_{D P_{i}}\right\| Y_{1} \| Y_{2}^{*}\right)$. Finally, $DP_i$ generates $MSG_2$ including $Y_1$, $M_3$, $M_4$, and $E_{X_{AV_{i}}}\left(ID_{DP_{i}} \|ID^{**}_{P_i}\right)$, then sends $MSG_2$ to $AV_i$.

$\bm{Step\ Au_2}$: $AV_i$ computes $y_{2}^{\prime}=M_{4} \oplus h\left(x_{A V_{i}}\left\|I D_{A V_{i}}\right\| Y_{1}\right)$ and $M^\prime_{3}=h\left(ID_{AV_{i}}\left\|Y_{1}\right\|x_{AV_{i}}\|y^\prime_{2}\right)$, then checks whether the equation $M^*_3=M_3$ holds. Turn to $\bm{Step\ Au_3}$ if $M^*_3=M_3$, otherwise terminate authentication.

$\bm{Step\ Au_3}$: $AV_i$ chooses a random number $y_3$ to calculate $Y_3=y_{3}P$ $Y_4=y_{3}Y_1$, $M_{5}=h\left(y_{2}^{\prime}\left\|Y_{3}\right\|x_{AV_{i}}\left\|ID_{AV_{i}}\right\|Y_{1}\right)$, $ADSK=h\left(ID_{AV_{i}}\left\|ID_{DP_{i}}\right\|Y_{3}\|y_{2}\right)$, and $PASK=h\left(ID^{**}_{P_{i}}\left\|ID_{AV_{i}}\right\|Y_{1}\|Y_{4}\right)$. $AV_i$ generates $MSG_3$ containing $Y_5$ and $M_5$, then sends $MSG_3$ to $DP_i$.

$\bm{Step\ Au_4}$: $DP_i$ calculates $M^*_{5}=h(y_{2}||Y_{3}||x^*_{AV_{i}}||ID^\prime_{AV_{i}}||Y_{1})$ and checks if the equation $M^*_5=M_5$ holds. Note that $DP_i$ and $AV_i$ complete mutual authentication if $M^*_5=M_5$, or else terminate authentication. Next, $DP_i$ computes $M_{6}=h(x_{P_{l}}^{\prime}||Y_{1}||Y_{2}^{*}||ID_{P_{i}}^{**}||ID_{AV_{i}}^{\prime} \|Y_{3})$ and $ADSK=h\left(ID_{AV_{i}}\left\|ID_{DP_{i}}\right\|Y_{3}\|y_{2}\right)$. Finally, $DP_i$ generates $MSG_4$, $M_6$, $Y_3$, and $E_{X_{P_{i}}}{(ID_{DP_i})}$, then sends $MSG_4$ to $DP_i$.

$\bm{Step\ Au_5}$: $P_i$ decrypts $E_{X_{P_{i}}}{(ID_{DP_i})}$ to obtain $ID_{DP_{i}}$ and calculates $M^*_{6}=h(x^*_{P_{l}}||Y_{1}||Y_{2}||ID^*_{P_{i}}||ID_{AV_{i}}||Y_{5})$. Note that $P_i$ and $DP_i$ finish mutual authentication if $M^*_6=M_6$, then turn to $\bm{Step\ Au_6}$, otherwise terminate authentication.

$\bm{Step\ Au_6}$: $DP_i$ calculates $Y^*_4=y_1Y_3$, $PDSK=h(ID^{*}_{P_{i}}||ID_{DP_{i}}||Y_{1}||Y_{2})$, and $PASK=h(ID^{*}_{P_{i}}||ID_{AV_{i}}||Y_{1}||Y^{*}_{4})$.

\noindent {3)\textbf{\ Ride Request}}

After $P_i$ finishes login and authentication, he / she sends a ride request including identity $ID^*_{P_{i}}$, destination $Des$, parking moment $T_{Parking}$, parking duration $t_{Parking}$, and timestamp $T_{1}$ to the adjacent pick-up point $DP_{i}$.

\noindent {4)\textbf{\ Matching}}

$DP_{i}$ chooses the $DP$ with the shortest passengers travel time as the best drop-off point (hereinafter defined as $ID_{DP_k}$) and the $PL$ with the smallest parking time cost as the optimal parking lot (hereinafter defined as $ID_{PL_i}$). The matching $AV$ (hereinafter defined as $ID_{AV_i}$) is selected based on the earliest available moment of $AVs$.

\noindent {5)\textbf{\ Pick Up Passenger \& Authentication}}

$DP_i$ sends $ID_{AV_i}$ to $P_i$ and sends the acknowledgment message $ACK$ to $AV_i$. $P_i$ walks to $DP_i$ waiting for $AV_i$. Additionally, $P_i$ sends $E_{PASK}(h(ID_{AV_{i}}), loc_{DP_k},ID_{P_i})$ to $AV_i$, which will pick up $P_i$ only if $h{(ID_{AV_{i}})}$ is valid.
\subsection{Drop Off \& Parking}
\noindent {1)\textbf{\  Parking Request}}

During taking $P_i$ to $loc_{DP_k}$, $AV_i$ generates ciphertext $MSG_{5}=\{E_{ADSK}\left(ID_{AV_{i}}\left\|T_{2}\right\|{Req}_{parking}\right), ID_{DP_{i}}\}$ via the session key $ADSK$, where $T_2$ and $Req_{parking}$ are the timestamp and parking request. Moreover, $AV_i$ sends $MSG_{5}$ to the adjacent $DP$ (hereinafter defined as $DP_j$), which forwards $MSG_5$ to $DP_i$. Unsurprisingly, $DP_i$ verifies the validity of $T_2$ and whether $ID_{AV_{i}}$ exists in the local database. If the above verification is successful, $DP_i$ sends the acknowledgment message $ACK$ and $MSG_6=E_{ADSK}(loc_{PL_i},T_3)$ to $DP_j$, which forwards $MSG_6$ to $AV_i$. Moreover, $DP_j$ selects the random string $Park_{AV_i}$ and signs it to generate the signature $Sig=x_{DP_j}h{(Park_{AV_i})}$. Subsequently, $DP_j$ sends $(Park_{AV_i},Sig)$ to $AV_i$. In response, $AV_i$ sends the confirmation message $Con_{AV_{i}}$ to $DP_k$, indicating that $AV_i$ is willing to park at $PL_i$. $DP_k$ generates the parking code $Code_{Parking}$ and signature ${Sig}^{*}=x_{DP_{k}}h{(Code_{Parking})}$. At last, $DP_k$ respectively sends $(Code_{Parking},Sig^{*})$ and $Code_{Parking}$ to $AV_i$ and $PL_i$. $Code_{Parking}$ is stored in the local database by $PL_i$.

\noindent {2)\textbf{\  Drop Off Passenger \& Authentication}}

 After arriving at $DP_k$, $AV_i$ sends $(Park_{AV_i},Sig)$ to $DP_k$, which verifies whether $e(Sig,P)=e(h(Park_{AV_i}),X_{DP_j})$ holds. If the above equation holds, $P_i$ can get off at $DP_k$.

\noindent {3)\textbf{\  Parking Check-in}}

$AV_i$ sends $(Code_{Parking},Sig^{*})$ to $PL_i$ after arriving at $PL_i$. $PL_i$ first retrieves whether $Code_{Parking}$ exists in the local database, $AV_i$ is prohibited to enter $PL_i$ in case nonexistent $Code_{Parking}$. Otherwise, $PL_i$ verifies the signature validity, that is, to check whether the equation $e\left({Sig}^{*},P\right)=e\left(h(Code_{Parking}),X_{DP_{k}}\right)$ holds. $AV_i$ can drive into $PL_i$ if the above equation is established, and then $PL_i$ deletes $Code_{Parking}$ from the local database.
\subsection{Update \& Revocation}
\noindent {1)\textbf{\  Passwords and Biometrics Update}}

Passengers can update their passwords and biometrics (hereinafter defined as $PBU$) locally. The detailed process is as follows.

$\bm{Step\ PBU_1}$: $P_i$ inputs identity $ID^*_{P_i}$, original passwords $PW^*_{P_i}$, biometrics $Bio^*_{P_i}$, and the random number $\theta_{P_i}$ to $SC$, which calculates biometric key $\alpha_{P_{i}}^{*}={KeyRet}\left(Bio^{*}_{P_{i}},\beta_{P_{i}}\right)$. Then $SC$  calculates $APW_{P_{i}}^{*}=h(PW_{P_{i}}^{*}||\alpha_{P_{i}}^{*}||\theta_{P_{i}})$, $x_{P_i}^{*}=A_{P_{i}}\oplus h\left(APW_{P_{i}}^{*} \| ID_{P_{i}}^{*}\right)$, and $B_{P_{i}}^{*}=h(h(ID_{P_{i}}^{*}||PW_{P_{i}}^{*}||\alpha_{p_{i}}^{*}\|x_{B_{i}}^{*})mod\ n)$. Afterwards, $SC$ judges whether the equation $B_{P_{i}}^{*}=B_{P_{i}}$ holds, terminate the update if $B_{P_{i}}^{*}=B_{P_{i}}$ isn't established. Otherwise, $SC$ calculates $A P W_{P_{i}}^{\prime *}=h\left(P W_{P_{i}}^{*}\left\|\alpha_{P_{i}}^{*}\right\| \theta_{P_{i}}^{\prime}\right)$ and $A_{p_{i}}^{\prime *}=h\left(A P W_{P_{i}}^{*} \| I D_{P_{i}}^{*}\right) \oplus x_{P_{i}}^{*}$, where $\theta_{P_{i}}^{\prime}$ is a random number chosen by $P_i$. Finally, $SC$ prompts $P_i$ to input the updated passwords $PW_{P_{i}}^{new}$ and updated biometrics $Bio_{P_{i}}^{new}$.

$\bm{Step\ PBU_2}$: $SC$ calculates the updated biometric key $\alpha_{P_{i}}^{new}$ and the updated public parameter $\beta_{P_{i}}^{new}$ (i.e. $\left(\alpha_{P_{i}}^{\text {new }},\beta_{P_{i}}^{new}\right)={KPGen{(Bio^{new}_{P_i})}}$ when $P_i$ inputs $PW^{new}_{P_i}$ and $Bio_{P_{i}}^{new}$. Next, $SC$ calculates $APW_{P_{i}}^{'new}=h(PW_{P_{i}}^{new}||\alpha_{P_{i}}^{new}||\theta_{P_{i}})$, $x_{P_{i}}^{new}=A_{P_{i}}^{\prime *} \oplus h(APW_{P_{i}}^{\prime new} ||ID_{P_{i}}^{*})$, $B_{P_{i}}^{new}=h(h(ID_{P_{i}}^{*}||PW_{P_{i}}^{new}||\alpha_{P_{i}}^{new}||x_{P_{i}}^{new}mod \ n)$, $APW_{P_{i}}^{new}=h(PW_{P_{i}}^{new}||\alpha_{P_{i}}^{new}||\theta_{P_{i}}^{\prime})$, and $A_{P_{i}}^{new}=x_{P_{i}}^{new}\oplus h(APW_{P_{i}}^{new}||ID_{P_{i}}^{*})$.

$\bm{Step\ PBU_3}$: $SC$ updates the parameter $\{A_{P_{i}},B_{P_{i}},\theta_{P_{i}}^{\prime},\beta_{P_{i}}, X_{DP_{i}},P,n,h()\}$ as $\{A_{P_{i}}^{new},B_{P_{i}}^{new},\theta_{P_{i}}^{\prime},\beta_{P_{i}}^{new},X_{DP_{i}},P,n,h()\}$.

\noindent {2)\textbf{\  Smart Card Revocation}}

The detail process of smart card revocation (hereinafter defined as $SCR$) is as follows.

$\bm{Step\ SCR_1}$: $P_i$ sends identity $ID^*_{P_i}$ and smart card revocation request to the adjacent $DP$, which retrieves whether $ID^*_{P_i}$ exists in the local database. If $ID^*_{P_i}$ doesn't exist, terminate the smart card revocation. Otherwise, $DP$ verifies if $L$ is less than or equal to 5. Only if $L$ isn't greater than 5, $DP$ calculates $x_{P_{i}}^{n}=h\left(ID_{P_{1}}^{*}\left\|x_{DP_{i}}\right\|T_{7}\right)$ and $A_{P_{i}}^{'n}=h\left(APW_{P_{i}}\|ID_{R}^{*}\right)\oplus x_{P_{i}}^{n}$, where $T_7$ is the timestamp.

$\bm{Step\ SCR_2}$: The adjacent $DP$ creates a new list $L^{new}$ to count the number of inputting wrong passwords. Moreover, the initial value and the threshold value of $L^{new}$ are 0 and 5. Then, the adjacent $DP$ updates $\left\{ID_{P_{i}},T_{0},L\right\}$ as $\left\{ID_{P_{i}}^{*},T_{7}, L^{new}\right\}$. Finally, $DP$ writes $\left\{A_{P_{t}}^{\prime n},X_{D P_{i}},P\right\}$ into the new smart card, which is sent to $P_i$.

$\bm{Step\ SCR_3}$: This step is the same as $\bm{Step\ PR_3}$ and we won't repeat it.

\section{Security Analysis}
\subsection{Formal Security Proof}
\textbf{Theorem 1.} Let $\mathcal{A}$ be the adversary of breaking AKA security in polynomial time $pt$, then the advantage of $\mathcal{A}$ breaking AKA security is estimated as follows
\begin{equation}
Adv_{\mathcal{A}}^{{\mathop{\rm Sec}\nolimits} LAVP}(pt) \le \frac{{Q_{hash}^2}}{{|hash|}} + 2(\frac{{{Q_{{\rm{S}}end}}}}{{|D|}} + Adv_{\mathcal{A}}^{ECDDH}(pt))
\nonumber
\end{equation}
Where ${Q_{hash}}$ and ${Q_{Send}}$ are the times of $hash$ queries and $Send$ queries, respectively. $|hash|$ and $|D|$ represent the space of the hash function and the size of the password dictionary, respectively. $Adv_{\mathcal{A}}^{ECDDH}$ is the advantage of breaking $ECDDH$ problem (\textbf{Definition 2.}).

\textbf{Proof.} Set $\Pr [Su{c_{G{m_i}}}]$ and $Ad{v_{G{m_i}}}$ be the probability and advantage of $\mathcal{A}$ winning games $\bm{G{m_i}(i = 0,1,2,3,4)}$, respectively. $Ad{v_{G{m_i}}} = \Pr [Su{c_{G{m_i}}}]$ and games $\bm{G{m_i}(i = 0,1,2,3,4)}$ are defined as follows.

\textbf{Game \bm{$Gm_0$}}: \bm{$Gm_0$} simulates real attacks with a random oracle model. $\mathcal{A}$ can challenge all queries. By \textbf{Definition 1.}, we have
\begin{equation}
Adv_{\mathcal{A}}^{{\mathop{\rm Sec}\nolimits} LAVP}(pt) = |2\Pr [Su{c_{G{m_0}}}] - 1|
\end{equation}

\textbf{Game \bm{$Gm_1$}}: In \bm{$Gm_1$}, $\mathcal{A}$ conducts eavesdropping attacks via $Execute(\Pi _{{P_i}}^i,\Pi _{D{P_i}}^j,\Pi _{A{V_i}}^k)$ queries. After \bm{$Gm_1$} is over, $\mathcal{A}$ performs a $Test$ query to determine whether the output of \bm{$Gm_1$} is a real session key or a random string. Since messages $MS{G_i}(i = 1,2,3,4)$ will not disclose the privacy information about the session key, the probability of winning the game will not be affected. So we have
\begin{equation}
\Pr [Su{c_{G{m_1}}}] = \Pr [Su{c_{G{m_0}}}]
\end{equation}

\textbf{Game \bm{$Gm_2$}}: $\mathcal{A}$ deceives a participant via \bm{$Gm_2$} to receive forged message. Additionally, $\mathcal{A}$ launches active attacks via $Send$ queries. $MS{G_i}(i = 1,2,3,4)$ are associated with random numbers and identities, so the collision probability in $Send$ queries can be ignored. Based on the birthday paradox, we can get
\begin{equation}
|\Pr [Su{c_{G{m_2}}}] - \Pr [Su{c_{G{m_1}}}]| \le \frac{{Q_{hash}^2}}{{|hash|}}
\end{equation}

\textbf{Game \bm{$Gm_3$}}: $\mathcal{A}$ simulates $CorruptSC$ queries via \bm{$Gm_3$}. $\mathcal{A}$ can guess low-entropy passwords by offline dictionary attacks, but SecLAVP restricts $\mathcal{A}$ to enter wrong passwords at most 5 times. Therefore
\begin{equation}
|\Pr [Su{c_{G{m_3}}}] - \Pr [Su{c_{G{m_2}}}]| \le \frac{{{Q_{Send}}}}{{|D|}}
\end{equation}

\textbf{Game \bm{$Gm_4$}}: $\mathcal{A}$ simulates $CorruptAV$ queries via \bm{$Gm_4$}. Let ${Y_1} = uP$ and ${Y_3} = vP$. $Adv_{\mathcal{A}}^{ECDDH}$ is the advantage of breaking $ECDDH$ problem. Given $uP$ and $vP$, $\mathcal{A}$ can distinguish between $uvP$ and a random number. Therefore
\begin{equation}
|\Pr [Su{c_{G{m_4}}}] - \Pr [Su{c_{G{m_3}}}]| \le Adv_{\mathcal{A}}^{ECDDH}(pt)
\end{equation}

After $\mathcal{A}$ runs all queries, the probability of guessing the correct bit $\ell$ is 0.5, so
\begin{equation}
\Pr [Su{c_{G{m_4}}}] = 1/2
\end{equation}

According to equations (1)(2)(6), we get
\begin{equation}
\begin{aligned}
\frac{1}{2}Adv_{\mathcal{A}}^{{\mathop{\rm Sec}\nolimits} LAVP}(pt)
 &= |\Pr [Su{c_{G{m_0}}}] - \frac{1}{2}|\\
 &= |\Pr [Su{c_{G{m_1}}}] - \Pr [Su{c_{G{m_4}}}]|
\end{aligned}
\end{equation}

Based on equations (3)-(5), we get
\begin{equation}
\begin{aligned}
|\Pr [Su{c_{G{m_1}}}] - \Pr [Su{c_{G{m_4}}}]|
 &\le |\Pr [Su{c_{G{m_1}}}] - \Pr [Su{c_{G{m_2}}}]|\\
 &+|\Pr [Su{c_{G{m_2}}}] - \Pr [Su{c_{G{m_4}}}]|\\
 &\le |\Pr [Su{c_{G{m_1}}}] - \Pr [Su{c_{G{m_2}}}]|\\
 &+ |\Pr [Su{c_{G{m_2}}}] - \Pr [Su{c_{G{m_3}}}]|\\
 &+ |\Pr [Su{c_{G{m_3}}}] - \Pr [Su{c_{G{m_4}}}]|\\
 &\le \frac{{Q_{hash}^2}}{{2|hash|}}+\frac{{{Q_{{\rm{S}}end}}}}{{|D|}}+Adv_{\mathcal{A}}^{ECDDH}(pt)
\end{aligned}
\end{equation}

Based on equations (7) and (8), we have
\begin{equation}
Adv_{\mathcal{A}}^{{\mathop{\rm Sec}\nolimits} LAVP}(pt) \le \frac{{Q_{hash}^2}}{{|hash|}} + 2(\frac{{{Q_{{\rm{S}}end}}}}{{|D|}} + Adv_{\mathcal{A}}^{ECDDH}(pt))
\end{equation}
\subsection{Formal Security Verification}
We apply AVISPA \cite{33} to formally verify the security of SecLAVP. AVISPA is a widely accepted protocol evaluation tool, which utilizes HLPSL (High Level Protocol Specification Language) to describe message interaction and entity roles. Moreover, AVISPA includes four backends, namely OFMC, CL-AtSe, SATMC, and TA4SP, where OFMC is widely used in verifying protocol security \cite{38}. The output of SUMMARY is `SAFE' if the protocol is secure, otherwise `UNSAFE' and the attack process are output.

\begin{figure}[htbp]
\setlength{\abovecaptionskip}{-0.1cm}
\setlength{\belowcaptionskip}{-0cm}
\centering
\includegraphics[scale=0.96]{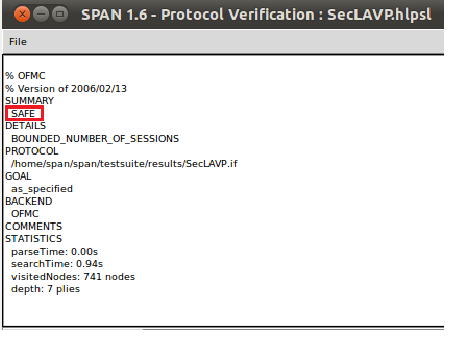}
\caption{Formal verification results of SecLAVP}
\end{figure}

We implement passenger registration, $AV$ registration, login, and authentication through HLPSL. The defined goals are to ensure the confidentiality of passwords and biometrics. In Login phase, $DP$ authenticates passenger via $M_1$. In authentication phase, $AV$ authenticates $DP$ via $M_2$. Similarly, $DP$ authenticates $AV$ via $M_3$ and passenger authenticates $DP$ via $M_4$. Intruders can not only eavesdrop and tamper with messages, but also act as intermediaries to intercept message. Fig. 5 shows the formal verification results of SecLAVP, which indicates that SecLAVP can resist man-in-the-middle attacks.

\subsection{Informal Security Analysis}
We conduct informal security analysis based on design goals $DG_i$. The strengths of SecLAVP are shown in Table \uppercase\expandafter{\romannumeral3}.
\begin{table}[htbp]\scriptsize
\setlength{\abovecaptionskip}{-0.05cm}
\setlength{\belowcaptionskip}{2pt}
 \centering
 \caption{Comparison of security}
 \label{table_time}
 \begin{spacing}{1.1}
 \begin{tabular}{|c|c|c|c|c|c|}
  \hline
   Security &Ref \cite{27} &Ref \cite{28} &Ref \cite{29} &Ref \cite{42} &SecLAVP\\
  \hline
  %\midrule
   $DG_1$ & $\times$ & \checkmark & \checkmark & \checkmark & \checkmark\\
  \hline
   $DG_2$ & \checkmark & \checkmark & \checkmark & \checkmark & \checkmark\\
  \hline
   $DG_3$ & \checkmark & \checkmark & \checkmark & \checkmark & \checkmark\\
  \hline
   $DG_4$ & $\times$ & $\times$ & \checkmark & \checkmark & \checkmark\\
  \hline
   $DG_5$ & $\times$ & \checkmark & \checkmark &$\times$ & \checkmark\\
  \hline
   $DG_6$ & \checkmark & \checkmark & \checkmark & \checkmark & \checkmark\\
  \hline
   $DG_7$ & \checkmark & \checkmark & \checkmark & \checkmark & \checkmark\\
  \hline
   $DG_8$ & \checkmark & $\times$ & \checkmark & \checkmark & \checkmark\\
  \hline
   $DG_9$ & $\times$ & $\times$ & $\times$ & \checkmark & \checkmark\\
  \hline
   $DG_{10}$ & \checkmark & \checkmark & \checkmark & \checkmark & \checkmark\\
  \hline
   $DG_{11}$ & $\times$ & $\times$ & $\times$ & \checkmark & \checkmark\\
  \hline
   $DG_{12}$ & $\times$ & $\times$ & $\times$ & $\times$  & \checkmark\\
  \hline
   $DG_{13}$ & $\times$ & \checkmark & $\times$ & $\times$ & \checkmark\\
  \hline
   $DG_{14}$ & $\times$ & $\times$ & $\times$ & $\times$ & \checkmark\\
  \hline
   $DG_{15}$ & \checkmark & $\times$ & $\times$ & $\times$ & \checkmark\\
  \hline
  %\bottomrule
 \end{tabular}
 \end{spacing}
\end{table}

1)\ \textbf{\bm{$D{G_1}$} Anonymity}: Attackers get the passenger's identity via $Y_1$ $Y_2$ and $M_2$, in which the calculation of ${Y_2}={x_{D{P_i}}}Y_1$ must obtain $DP's$ private key $x_{{DP}_i}$. However, $x_{{DP}_i}$ is only known to $DP_i$. So SecLAVP satisfies anonymity.

2)\ \textbf{\bm{$D{G_2}$} Mutual Authentication}: The $SC$ authenticates the passenger via $B^*_{P_i}$. Additionally, the $DP$ and the $AV$ achieve mutual authentication by $M_3$ and $M_5$. The $DP$ and the passenger achieve mutual authentication by $M_1$ and $M_6$.

3)\ \textbf{\bm{$D{G_3}$} AKA Security}: See part $A$ of this Section for details.

4)\ \textbf{\bm{$D{G_4}$} Forward Security}: In polynomial time, due to $ECDDH$ problem, attackers can't calculate ${Y_2}={y_1}{X_{D{P_i}}}$ based on $Y_1$ and $P$. Therefore, the session key between the passenger and the $DP$ won't be compromised. Furthermore, due to the confidentiality of $y_2$ and $y_3$, the session key between the passenger and the $AV$ as well as the session key between the $DP$ and the $AV$ won't be compromised.

5)\ \textbf{\bm{$D{G_5}$} Update}: Passwords and biometrics update can effectively resist passwords guessing and biometrics leakage.

6)\ \textbf{\bm{$D{G_6}$} No Passwords Verified Table}: Secret parameters related to passwords aren't disclosed because $DP$ only store $ID_{P_i}$, $T_0$, and $L$ in the registration phase.

7)\ \textbf{\bm{$D{G_7}$} Passwords Security}: A random number $\theta _{{P_i}}^{'}$ selected by the passenger prevents $DP$ from obtaining passwords.

8)\ \textbf{\bm{$D{G_8}$} Anti Smart Card Loss}: Supposing that attackers can obtain the parameters $\{ {A_{{P_i}}},{B_{{P_i}}},\theta _{{P_i}}^{'},{\beta _{{P_i}}},{X_{D{P_i}}},P,n,h()\}$ and the biometrics when $SC$ is lost. Since passwords are obfuscated by $B_{P_i}$ and the hash function is collision resistance, attackers can't infer passwords based on the known information.

9)\ \textbf{\bm{$D{G_9}$} Internal Attacks Resistance}: In SecLAVP, internal attacks \cite{39} means that $DP$ can guess passwords. Even though $DP$ obtains $ID_{P_i}$, $APW_{P_i}$, $Bio_{P_i}$, and $\{{A_{{P_i}}},{B_{{P_i}}},\theta _{{P_i}}^{'},{\beta _{{P_i}}},{X_{D{P_i}}},P,n,h()\}$, $DP$ can't calculate $AP{W_{{P_i}}}= h(P{W_{{P_i}}}||{\alpha _{{P_i}}}||{\theta _{{P_i}}})$. Consequently, $DP$ can't confirm whether the guessed passwords are correct.

10)\ \textbf{\bm{$D{G_{10}}$} Replay Attacks Resistance}: $DP$ verifies the validity of message via timestamp to resist replay attacks.

11)\ \textbf{\bm{$D{G_{11}}$} Offline Dictionary Guessing Attacks Resistance}: Supposing that the attacker applies the scanner and side channel attacks to steal the biometrics and parameters$\{{A_{{P_i}}},{B_{{P_i}}},\theta_{{P_i}}^{'},{\beta_{{P_i}}},{X_{D{P_i}}},P,n,h()\}$ stored in the smart card. Attackers guess the passenger's identity and passwords as $\{i{d_{{P_i}}},p{w_{{P_i}}}\}$. Next, attackers calculates $\alpha _{_{{P_i}}}^{*} = KeyRet{(Bi{o_{P_i}},{\beta _{P_i}})}$, $ap{w_{{P_i}}} = h(p{w_{{P_i}}}||\alpha _{_{{P_i}}|}^*|\theta _{_{{P_i}}}^*)$, $x_{_{{P_i}}}^* = h(ap{w_{{P_i}}}||i{d_{{P_i}}}) \oplus {A_{{P_i}}}$, and $B_{_{{P_i}}}^* = h(i{d_{{P_i}}}||p{w_{{P_i}}}||\alpha _{_{{P_i}}}^*||x_{_{{P_i}}}^*)\bmod n$. Attackers judges whether the identity and passwords are guessed correctly according to the equation $B_{_{{P_i}}}^{*} = {B_{{P_i}}}$. There are ${2^{32}}$ $\{i{d_{{P_i}}},p{w_{{P_i}}}\}$ satisfying $B_{_{{P_i}}}^{*} = {B_{{P_i}}}$ \cite{40}, but attackers can only guess 5 times at most. So the probability of guessing passwords correctly is ${5\mathord{\left/{\vphantom {5 {{2^{32}}}}} \right.\kern-\nulldelimiterspace}{{2^{32}}}}\approx 1.16 \times{10^{ - 9}}$, which is negligible.

12)\ \textbf{\bm{$D{G_{12}}$} Multiple-reservation Resistance}: $DP$ generates a one-time parking code for parking check-in and stores it in the local database. Additionally, $DP$ will delete the historical parking code once $AV$ enters the $PL$, which is convenient for $DP$ to check multiple-reservation in next round.

13)\ \textbf{\bm{$D{G_{13}}$} Efficiency}: SecLAVP is efficient under the premise of ensuring security, which is mainly reflected in communication overheads (see Section \uppercase\expandafter{\romannumeral7}. A), computational overheads (see Section \uppercase\expandafter{\romannumeral7}. B), and scheduling performance (see Section \uppercase\expandafter{\romannumeral7}. C).

14)\ \textbf{\bm{$D{G_{14}}$} Formal Security Proof}: See part $A$ of this Section for details.

15)\ \textbf{\bm{$D{G_{15}}$} Formal Security Verification}: See part $B$ of this Section for details.

\section{Performance Evaluation}
\subsection{Communication Overheads}
Communication overheads generated by SecLAVP are mainly message (${MSG_{1}\sim MSG_{4}}$). The comparison of communication overheads is shown in Table \uppercase\expandafter{\romannumeral4}. The identity, random number, hash function and the key of $ECC$ are 128$bits$, 128$bits$, 160$bits$ and 160$bits$.

\begin{table}[htbp]\scriptsize
\setlength{\abovecaptionskip}{-0.05cm}
\setlength{\belowcaptionskip}{2pt}
 \centering
 \caption{Comparison of communication overheads (in bits)}
 \label{table_time}
 \begin{spacing}{1.06}
 \begin{tabular}{|c|c|c|c|c|c|}
 \hline
   Scheme &$MSG_1$ &$MSG_2$ &$MSG_3$ &$MSG_4$ &Total\\
  \hline
   Ref \cite{27} &992 &960 &448 &864 &3264\\
  \hline
   Ref \cite{28} &763 &576 &288 &576 &2203\\
  \hline
   Ref \cite{29} &864 &960 &288 &576 &2688\\
  \hline
  Ref \cite{42}  &768 &1344 &960 &480 &3552\\
  \hline
   SecLAVP &640 &736 &320 &448 &2144\\
  \hline
 \end{tabular}
 \end{spacing}
\end{table}

Fig. 6 is a comparison diagram about communication overheads. $Y_1$, $M_1$, $M_2$, and $PID_{AV_i}$ are all 160$bits$. Therefore, communication overheads of $MS{G_1}=\{ {Y_1},{M_1},{M_2},PI{D_{A{V_i}}}\}$ is $160 \times 4 = 640bits$. $M_3$ and $M_4$ are all 160$bits$. $I{D_{D{P_i}}}$ and $I{D_{A{V_i}}}$ are all 128$bits$. Thus, communication overheads of $MS{G_2} = \{ {Y_1},{M_3},{M_4},{E_{{X_{A{V_i}}}}}(I{D_{D{P_i}}},I{D_{A{V_i}}})\}$ is $160 \times 3 + 128 \times 2 = 736bits$.  Similarly, $M_5$, $M_6$, and $Y_3$  are all 160$bits$. Hence, communication overheads of $MS{G_3}=\{ {Y_3},{M_5}\}$ and $MS{G_4} = \{ {M_6},{Y_3},{E_{{X_{A{V_i}}}}}(I{D_{D{P_i}}})\}$ are respectively $160 \times 2 = 320bits$ and $160 \times 2 + 128 = 448bits$. Communication overheads have been reduced by 23.92$\%$ on average.

\begin{figure}[htbp]
\setlength{\abovecaptionskip}{-0.1cm}
\setlength{\belowcaptionskip}{-0cm}
\centering
\includegraphics[scale=0.32]{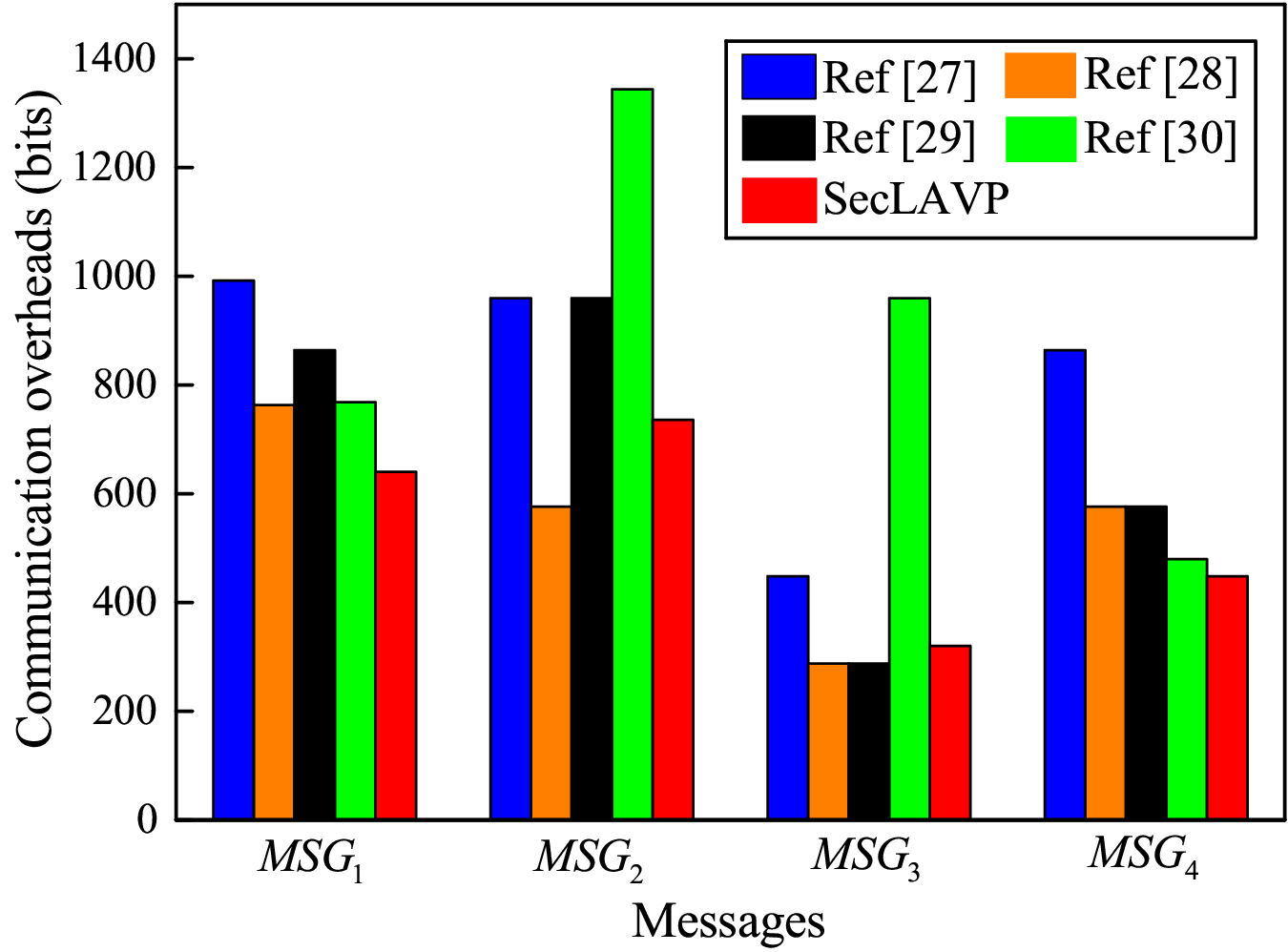}
\caption{Communication overheads comparison}
\end{figure}

\subsection{Computational Overheads}

Table \uppercase\expandafter{\romannumeral5}\ shows the comparison of computational overheads. Provided that $T_{ECM}$ and $T_{ECA}$ are multiplication operations and addition operations on $ECC$. Moreover, $T_H$ and $T_E$ are hash operations and encryption (or decryption) operations. A mobile phone with a memory of 3GB and a frequency of 2.45GHz simulates the passenger. Furthermore, the laptop computer with 16GB memory and a frequency of 3.1GHz emulates the autonomous vehicle. Besides, the desktop computer with 32GB memory and a frequency of 4.0 GHz serves as the drop-off/pick-up point \cite{29}. For the mobile phone, $T_{ECM}$, $T_{ECA}$, $T_H$, and $T_E$ take 0.537ms, 0.601ms, 11.260ms, and 13.434ms. Concerning the laptop computer, $T_{ECM}$, $T_{ECA}$, $T_H$, and $T_E$ take 0.050ms, 0.051ms, 0.728ms, and 1.587ms. As for the desktop computer, $T_{ECM}$, $T_{ECA}$, $T_H$, and $T_E$ take 0.041ms, 0.045ms, 0.483ms, and 0.978ms.
\begin{table*}[htbp]\scriptsize
\setlength{\abovecaptionskip}{-0.05cm}
\setlength{\belowcaptionskip}{2pt}
 \centering
 \caption{Comparison of the computational overheads(in millseconds)}
 \label{table_time}
 \begin{spacing}{1.15}
 \begin{tabular}{|c|c|c|c|c|}
  \hline
   Scheme &Passenger &$DP$ &$AV$ &Total\\
  \hline
   Ref \cite{27} &$7{T_H} + 2{T_E}$ &$8{T_H} + 5{T_E}$ &$4{T_H} + {T_E}$ &118.941\\
  \hline
   Ref \cite{28} &$7{T_H} + 2{T_E} + 2{T_{ECM}}$ &$8{T_H} + 4{T_E} + {T_{ECM}}$ &$4{T_H} + 2{T_E}$ &120.665\\
  \hline
   Ref \cite{29} &$9{T_H} + {T_E} + 5{T_{ECM}} + {T_{ECA}}$ &$12{T_H} + 2{T_E} + 5{T_{ECM}} + {T_{ECA}}$ &$4{T_H} + {T_E}$ &130.561\\
  \hline
   Ref \cite{42} &$10{T_H}$ &$5{T_H}$ &$11{T_H}$ &123.023\\
  \hline
   SecLAVP &$9{T_H} + {T_E} + 3{T_{ECM}}$ &$11{T_H} + 2{T_E} + {T_{ECM}}$ &$5{T_H} + {T_E} + 2{T_{ECM}}$ &129.022\\
  \hline
 \end{tabular}
  \end{spacing}
\end{table*}

\begin{figure*}
\setlength{\abovecaptionskip}{-0.05cm}
\setlength{\belowcaptionskip}{0cm}
\subfigure[Computational overheads comparison for passengers]{
\begin{minipage}[t]{0.343\linewidth}
\includegraphics[width=2.33in]{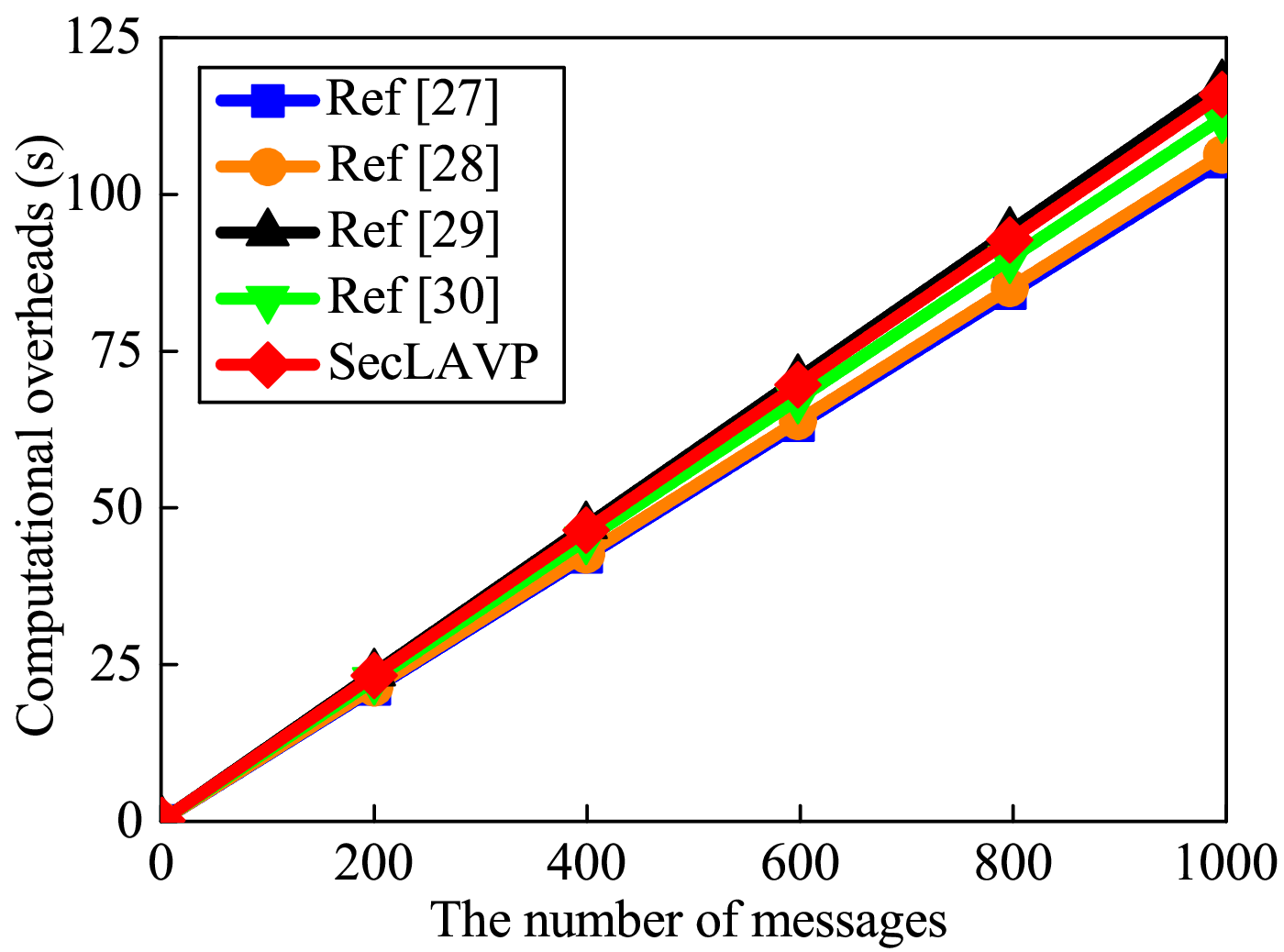}
%\caption{fig1}
\end{minipage}%
}%
\subfigure[Computational overheads comparison for $DPs$]{
\begin{minipage}[t]{0.33\linewidth}
\centering
\includegraphics[width=2.3in]{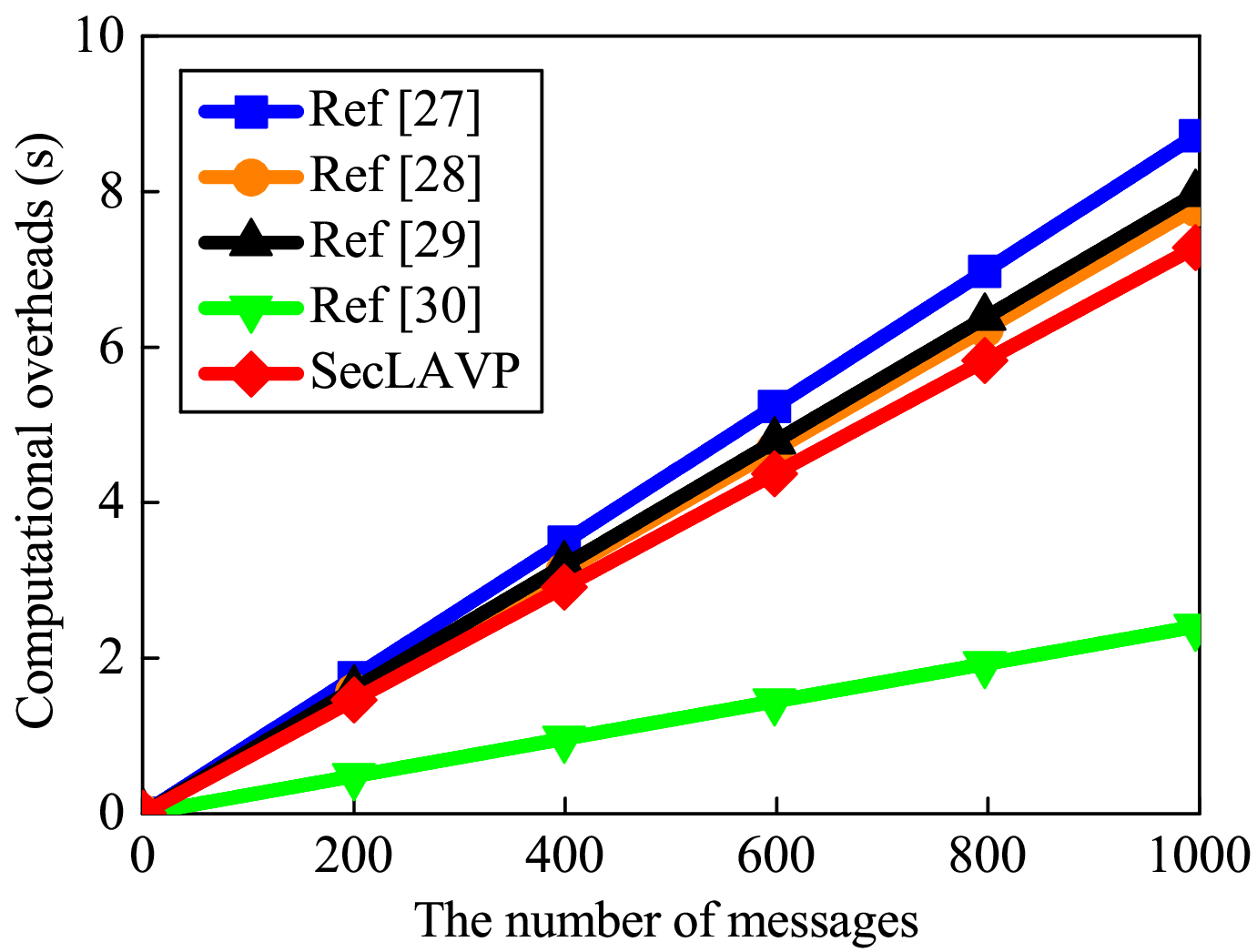}
%\caption{fig2}
\end{minipage}%
}%
\subfigure[Computational overheads comparison for $AVs$]{
\begin{minipage}[t]{0.345\linewidth}
\centering
\includegraphics[width=2.26in]{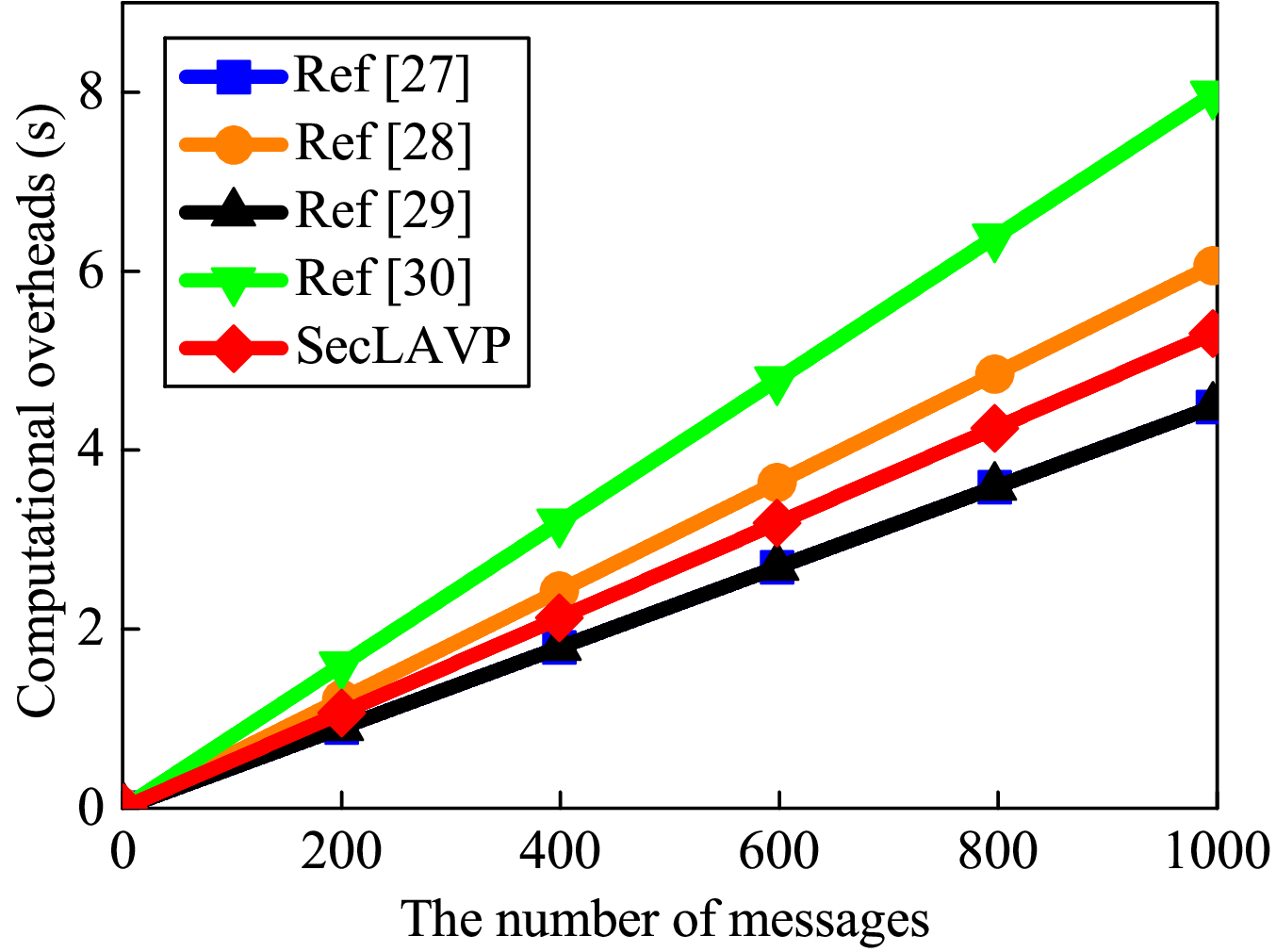}
%\caption{fig2}
\end{minipage}
}%
\centering
\caption{Computational overheads comparison}
\end{figure*}

In SecLAVP, the passenger calculates $APW_{{P_i}}^{*}$, $x_{_{{P_i}}}^{*}$, $B_{{P_i}}^{*}$, ${M_1}$, ${M_2}$, $PI{D_{A{V_i}}}$, $M_6^*$, $PDSK$, and $PASK$ consuming $9T_H$, then calculates $Y_1$, $Y_2$, and $Y_4^{*}$ consuming $3T_{ECM}$. Furthermore, the passenger obtains the identity of $DP_i$ consuming $T_E$. As a result, the passenger's computational overheads are $9{T_H} + 3{T_{ECM}}+{T_E}$.  Similarly, computational overheads of $DPs$ and $AVs$ are respectively $11{T_H}+{T_{ECM}}+2{T_E}$ and $5{T_H} + 2{T_{ECM}} + {T_E}$.

Computational overheads comparison of passengers is presented in Fig. 7\ (a), which indicates computational overheads augment with the quantity of messages. More importantly, compared with the reference \cite{29}, computational overheads of passengers in SecLAVP are preponderant with the increment of messages' amount. Compared with the solution \cite{27}, SecLAVP introduces $Y_1$ and $Y_2$ to defend attackers against guessing passwords via $M_1$, resulting in greater computational overheads. What's more, $B_{P_i}$ is applied to fuzzy processing passwords to resist smart card loss existing in the scheme \cite{28}. SecLAVP has greater computational overheads than the reference \cite{42}, because biometric extraction is applied to prevent biometrics leakage and biometrics update, which aren't included in \cite{42}.

Fig. 7\ (b) shows computational overheads comparison of $DPs$. Simulation results manifest computational overheads of $DPs$ are linearly related to the number of messages. Along with the growth of messages, SecLAVP is superior to schemes \cite{27,28,29}. SecLAVP takes 7.31s for 1000 messages, but schemes \cite{27,28,29} take respectively 8.754s, 7.817s, and 8.002s. SecLAVP enhances the security level by authenticating passengers through $M_1$ in the login, resulting in larger computational overheads than that of the scheme \cite{42}.

Fig. 7\ (c) compares computational overheads of $AVs$, which demonstrates computational overheads of $AVs$ increase linearly with the number of messages. Compared with references \cite{28,42}, computational overheads of $AVs$ in SecLAVP have more significant advantages with the increase of messages. Additionally, SecLAVP introduces $y_2$ and $Y_4$ to support AKA security, which isn't  included in \cite{27}. Even if computational overheads of $AVs$ in SecLAVP are greater than that of the scheme \cite{29}, computational overheads of both passengers and $DPs$ are less than that of the scheme \cite{29}.

\subsection{Scheduling Performance}
The scheduling performance is simulated on Opportunistic Network Environment (ONE) \cite{41}. We deploy 6 $PLs$ and 15 $DPs$ in the network, each $PLs$ containing 100 parking spaces \cite{10}. Additionally, the duration of $AV$ requesting for drop-off and parking duration are set 3200s and 7200s respectively. We compare the scheduling performance between LAVP regardless of security \cite{10} (hereinafter referred to as LAVP-RS) and SecLAVP including AKA, authentication in the pick-up phase, authentication in the drop-off phase, and parking check-in.

The scheduling performance is evaluated from the following three aspects:
\begin{itemize}
\item \textbf{Waiting duration for parking spaces}: The time interval between arriving at the parking lot and obtaining available the parking space (hereinafter referred to as waiting duration).
\item \textbf{Passenger travel time}: The travel duration from the pick-up point to the drop-off point, plus the walking duration from the drop-off point to the destination.
\item \textbf{\bm{$AV$} travel time}: The driving time from the pick-up point to the drop-off point, plus that from the drop-off point to the parking lot.
\end{itemize}

\noindent{1)\ \textbf{Influence of \bm{$AV$} Density}}

Table \uppercase\expandafter{\romannumeral6}\ shows the influence of $AV$ density on scheduling performance. In LAVP-RS and SecLAVP, waiting duration increases linearly with the quantity of $AVs$, whereas SecLAVP supports AKA for secure communication. Moreover, parking check-in guarantees that $AV$ drives to the matching parking lot. In the context of 300, 400, 500 and 600 $AVs$, the waiting duration of SecLAVP is slightly longer than or equal to that of LAVP-RS, but SecLAVP has certain advantages as $AVs$ rise to 700.

\begin{table}[htbp]\scriptsize
\setlength{\abovecaptionskip}{-0.05cm}
\setlength{\belowcaptionskip}{2pt}
 \centering
 \caption{The influence of AV density on scheduling performance(in seconds)}
 \begin{spacing}{1.15}
\begin{tabular}{|c|c|c|c|c|c|c|}
\hline
\multicolumn{1}{|c|}{\multirow{2}{*}{AV}} & \multicolumn{2}{c|}{Waiting duration}                            & \multicolumn{2}{c|}{Passenger travel time}      &\multicolumn{2}{c|}{$AV$ travel time}                             \\ \cline{2-7}
\multicolumn{1}{|c|}{}                    & \multicolumn{1}{c|}{LAVP-RS} & \multicolumn{1}{c|}{SecLAVP} & \multicolumn{1}{c|}{LAVP-RS} & \multicolumn{1}{c|}{SecLAVP} & \multicolumn{1}{c|}{LAVP-RS} & \multicolumn{1}{c|}{SecLAVP} \\ \hline
\multicolumn{1}{|c|}{300}                 & \multicolumn{1}{c|}{0}  & \multicolumn{1}{c|}{0.045}   & \multicolumn{1}{c|}{805.488}  & \multicolumn{1}{c|}{805.977}   & \multicolumn{1}{c|}{558.207}  & \multicolumn{1}{c|}{559.398}   \\ \hline
\multicolumn{1}{|c|}{400}      & \multicolumn{1}{c|}{22.525} & \multicolumn{1}{c|}{27.309}  & \multicolumn{1}{c|}{792.358}   & \multicolumn{1}{c|}{793.431}  & \multicolumn{1}{c|}{551.956}   & \multicolumn{1}{c|}{553.248}   \\ \hline
\multicolumn{1}{|c|}{500}                 & \multicolumn{1}{c|}{236.112}  & \multicolumn{1}{c|}{235.873}   & \multicolumn{1}{c|}{791.949}  & \multicolumn{1}{c|}{790.581}   & \multicolumn{1}{c|}{559.484}  & \multicolumn{1}{c|}{558.348}   \\ \hline
\multicolumn{1}{|c|}{600}                 & \multicolumn{1}{c|}{495.557}  & \multicolumn{1}{c|}{495.565}   & \multicolumn{1}{c|}{781.180}  & \multicolumn{1}{c|}{786.130}   & \multicolumn{1}{c|}{564.938}  & \multicolumn{1}{c|}{564.731}   \\ \hline
\multicolumn{1}{|c|}{700}                 & \multicolumn{1}{c|}{1076.217}  & \multicolumn{1}{c|}{1061.276}   & \multicolumn{1}{c|}{775.940}  & \multicolumn{1}{c|}{773.957}   & \multicolumn{1}{c|}{572.766}  & \multicolumn{1}{c|}{572.865}   \\ \hline
\end{tabular}
\end{spacing}
\end{table}

The passenger travel time decreases with the increase of $AV$. Compared with LAVP-RS, that of SecLAVP is slightly longer. Nonetheless, SecLAVP supports authentication in the pick-up phase and authentication in the drop-off phase, to prevent $AV$ from picking up the wrong passenger or casually dropping off a legitimate passenger.

In light of the fact that $AV$ travel duration is more affected by destination location, pick-up/drop-off point location, and parking lot location, the $AV$ density has a negligible effect on $AV$ travel duration.

\noindent{2)\ \textbf{Influence of \bm{$DP$} Density}}

Table \uppercase\expandafter{\romannumeral7}\ shows the influence of $DP$ density on scheduling performance. Under the condition of 4 $DPs$, LAVP-RS causes the parking hotspot in which $AVs$ are centrally allocated to a parking lot, resulting in a long waiting duration. In SecLAVP, the delay of parking check-in is so small that the scheduling performance isn't affected. Additionally, a small delay may lead $DPs$ to receive more reservations, which allows $DPs$ to make a better matching decision, resulting to less waiting duration. With the increase of $DPs$, although the waiting duration of SecLAVP is slightly longer than that of LAVP-RS, AKA and authentication at various phase included in SecLAVP enhance the security of reservation services. So SecLAVP has more strengths from the overall perspective.

\begin{table}[htbp]\scriptsize
\setlength{\abovecaptionskip}{-0.05cm}
\setlength{\belowcaptionskip}{2pt}
 \centering
 \caption{The influence of DP density on scheduling performance(in seconds)}
 \begin{spacing}{1.15}
\begin{tabular}{|c|c|c|c|c|c|c|}
\hline
\multicolumn{1}{|c|}{\multirow{2}{*}{DP}} & \multicolumn{2}{c|}{Waiting duration}                            & \multicolumn{2}{c|}{Passenger travel time}      &\multicolumn{2}{c|}{$AV$ travel time}                             \\ \cline{2-7}
\multicolumn{1}{|c|}{}                    & \multicolumn{1}{c|}{LAVP-RS} & \multicolumn{1}{c|}{SecLAVP} & \multicolumn{1}{c|}{LAVP-RS} & \multicolumn{1}{c|}{SecLAVP} & \multicolumn{1}{c|}{LAVP-RS} & \multicolumn{1}{c|}{SecLAVP} \\ \hline
\multicolumn{1}{|c|}{4}                 & \multicolumn{1}{c|}{778.8761}  & \multicolumn{1}{c|}{760.8562}   & \multicolumn{1}{c|}{1578.5951}  & \multicolumn{1}{c|}{1581.6658}   & \multicolumn{1}{c|}{961.5527}  & \multicolumn{1}{c|}{965.4355}   \\ \hline
\multicolumn{1}{|c|}{10}      & \multicolumn{1}{c|}{173.1871} & \multicolumn{1}{c|}{202.0649}  & \multicolumn{1}{c|}{1130.7221}   & \multicolumn{1}{c|}{1128.435}  & \multicolumn{1}{c|}{721.5275}   & \multicolumn{1}{c|}{718.7321}   \\ \hline
\multicolumn{1}{|c|}{15}       & \multicolumn{1}{c|}{0}       & \multicolumn{1}{c|}{0.045}  & \multicolumn{1}{c|}{805.488}   & \multicolumn{1}{c|}{805.977}  & \multicolumn{1}{c|}{558.207}   & \multicolumn{1}{c|}{559.3975}    \\ \hline
\end{tabular}
\end{spacing}
\end{table}

With the increase of $DP$, the optional drop-off points near the destination also increase, resulting in a decrease in passenger travel time. Besides, under the condition of the same $DP$ density, the passenger travel time of SecLAVP is almost equal to that of LAVP-RS .

The $AV$ travel duration decreases with the increase of $DPs$. Under the condition of 4 $DPs$, the $AV$ travel duration of SecLAVP is slightly longer than that of LAVP-RS. However, along with the amount of $DP$ increasing, the gap between LAVP-RS and SecLAVP gradually reduces and tends to zero.

\section{Conclusion}
In this paper, we propose a secure reservation scheme with three-factor AKA for LAVP, which generates the session key for secure communication. We then prove its security from formal security proof and formal security verification. Additionally, security analysis manifests that SecLAVP satisfies design goals ($DG_1-$$DG_{15}$). Compared with state-of-the-art AKA protocols, communication overheads have been reduced by 23.92\% on average. More importantly, the scheduling performance of SecLAVP is almost unaffected. To sum up, SecLAVP provides a theoretical foundation and guideline to construct a reservation scheme with a good balances security and availability for LAVP.
\ifCLASSOPTIONcaptionsoff
  \newpage
\fi

%\begin{thebibliography}{1}

\bibliographystyle{IEEEtran}
\bibliography{refer}
%\cite {1,2,3,4,5,6,7}

%\end{thebibliography}

%\begin{thebibliography}

%\end{thebibliography}

% insert where needed to balance the two columns on the last page with
% biographies
%\newpage

% if you will not have a photo at all:

\end{document}